\documentclass{IEEEtran}
\usepackage{bbm}
\usepackage{amssymb,amsmath,amsthm}
\usepackage{graphicx,epstopdf,psfrag,url,cite,xcolor,multirow,float}
\usepackage{booktabs}
\usepackage{caption-patch}
\usepackage[small,bf]{caption}
\usepackage{multirow}
\usepackage{makecell}
\usepackage{subcaption}
\usepackage[font=footnotesize]{subfig}
\usepackage[margin=0.7in]{geometry}

\usepackage{algorithm}
\usepackage{algorithmic}
\usepackage[super]{nth}
\usepackage{array}
\usepackage{pifont}
\usepackage{subfig}
\newcolumntype{C}[1]{>{\centering\let\newline\\\arraybackslash\hspace{0pt}}m{#1}}

\newtheoremstyle{mythm}% name
{\topsep}   % Space above1
{\topsep}   % Space below1
{\itshape}      % Body font
{0pt}       % Indent amount2
{\bfseries} % Theorem head font
{:}         % Punctuation after theorem head
{5pt plus 1pt minus 1pt}    % Space after theorem head3
{\thmname{#1}\thmnumber{ #2}\thmnote{ (#3)}}
\theoremstyle{mythm}

\newtheorem{proposition}{Proposition}

\newcommand{\bm}[1]{\boldsymbol{#1}}

\IEEEoverridecommandlockouts

\begin{document}

\title{Lyapunov-guided Deep Reinforcement Learning for Semantic-aware AoI Minimization in UAV-assisted Wireless Networks}

\author{Yusi Long, Shimin Gong, Sumei Sun, Gary Lee, Lanhua Li, Dusit Niyato

\thanks{Yusi Long, Shimin Gong, and Lanhua Li are with the School of Intelligent Systems Engineering, Shenzhen Campus of Sun Yat-sen University, Shenzhen 518000, China and Guangdong Provincial Key Laboratory of Fire Science and Intelligent Emergency Technology, China  (e-mail: longys@mail2.sysu.edu.cn, \{gongshm5, lilh65\}@mail.sysu.edu.cn).
Sumei Sun, and Gary Lee are with the Institute for Infocomm Research (I2R), A*STAR, Singapore (e-mail:\{sunsm, gary\_lee\}@i2r.a-star.edu.sg).
Dusit Niyato is with the College of Computing and Data Science, Nanyang Technological University, Singapore (e-mail: dniyato@ntu.edu.sg).
%\emph{(Corresponding author: Shimin Gong).}
}
}
\maketitle
\thispagestyle{empty}

\begin{abstract}
This paper investigates an unmanned aerial vehicle (UAV)-assisted semantic network where the ground users (GUs) periodically capture and upload the sensing  information to a base station (BS) via UAVs' relaying.
Both the GUs and the UAVs can extract semantic information from large-size raw data and transmit it to the BS for recovery.
Smaller-size semantic information reduces latency and improves information freshness, while larger-size semantic information enables more accurate data reconstruction at the BS, preserving the value of original information.
We introduce a novel semantic-aware age-of-information  (SAoI) metric to capture both information freshness and semantic importance, and then formulate a time-averaged SAoI minimization problem by jointly optimizing the UAV-GU association, the semantic extraction, and the UAVs' trajectories.
%To solve this computationally intensive problem,
We decouple the original problem into a series of subproblems via the Lyapunov framework and then use hierarchical deep reinforcement learning (DRL) to solve each subproblem.
Specifically, the UAV-GU association is determined by DRL, followed by the optimization module updating the semantic extraction strategy and UAVs' deployment.
Simulation results show that the hierarchical structure improves learning efficiency.
Moreover, it achieves low AoI through semantic extraction while ensuring minimal loss of original information, outperforming the existing baselines.
\end{abstract}
\begin{IEEEkeywords}
UAV, semantic communication, DRL, trajectory planning, Lyapunov optimization.
\end{IEEEkeywords}
\section{Introduction}
Future Internet of Things (IoT) has been extensively
used to collect real-time data from the surrounding environment for smart agriculture, autonomous driving, localization, and so on~\cite{AoI-survey, AoI-survey-2}.
In latency-sensitive applications, the freshness of data is of paramount importance.
To support these real-time applications, the sensing data generated by the IoT devices is expected to be transmitted to the destination as fresh as possible.
The age-of-information (AoI) as a metric of information freshness was proposed in~\cite{2017Aoi-origin, 2022IRS-aoi}, which is defined as the time elapsed since the most recent data update event, i.e., the recently received data has a smaller AoI value.
A smaller AoI signifies that the sensing information is fresher, providing more accurate feedback of the present state.
Conversely, a larger AoI suggests a longer delay since the sensing data was generated, which may result in discrepancies with the current environmental states.
Delayed information could potentially lead to erroneous control decisions or even catastrophic outcomes~\cite{aimin1,aimin22}.

The unmanned aerial vehicle (UAV)-assisted wireless network is considered a promising solution for real-time data transmission across expansive geographical areas due to the low deployment cost and high flexibility~\cite{2024Abs-UAV-real-time-Xi, 2024Abs-UAV-realtime-Liu}.
The UAVs can leverage Line-of-Sight (LoS) links to serve the ground users (GUs) better~\cite{2024-uav-aoi-Emami, 2024-uav-aoi-Liu}.
%The UAVs' trajectory planning ensures unobstructed communication between the UAVs and the GUs~\cite{2024-uav-aoi-Emami, 2024-uav-aoi-Liu}.
However, the UAVs' mobility causes dynamic and unstable channel conditions. This leads to signal attenuation and communication disruptions, which compromises the reliability of data transmission.
Furthermore, the channel competition between the UAVs and the GUs  exacerbates transmission uncertainty.
When the GUs are densely deployed, the intensified contention for spectral resources between the UAVs and the GUs leads to channel congestion and increased interference.
This not only increases the risk of data loss but also affects information updates.
Therefore, it is crucial to adapt the UAV-GU association strategy along with coordinating all UAVs' trajectories to keep the information fresh.
Previous studies on UAV-assisted AoI minimization mainly focused on conventional bit-oriented communication to guarantee reliable information delivery~\cite{2024-UAV-AoI-Pan, 2024-uav-aoi-Gong}.
Nonetheless, it is inefficient during the transmission of large-size data, as processing substantial amounts of redundant information increases transmission latency.
%Nonetheless, the intensive and repetitive data from IoT devices pose a challenge for efficient communication.

A new paradigm called semantic communication has recently been proposed to shift the focus of information delivery to the semantic level by  prioritizing context-aware information importance~\cite{2021-semantic-Xie,2024-semantic-uav-Liew}.
This technology allows essential content, namely semantic information, to be extracted from bit information at the transmitter and delivered to the receiver, which significantly reduces the amount of data that needs to be transmitted and decreases the communication overhead.
It prioritizes the transmission of critical information, enabling more effective and timely decision-making.
When semantic information is successfully received and recovered by the receiver, we can still ensure an accurate understanding of the sensing data and thus maintain the value of information.
The information value may be compromised if the original information cannot be exactly recovered from the received semantic information.
Intuitively, smaller-size semantic information requires shorter transmission delay.
This means that the BS receives the update information more quickly, resulting in a lower AoI.
However, it may increase the difficulty of information recovery at the receiver, potentially causing information loss and distortion, which compromises the integrity or value of the information.
Therefore, it is crucial to optimize the amount of extracted semantic information.
Formally, we can define the information value as the ratio of data that can be successfully recovered from the original data~\cite{2023-semantic-Li}.
A balance between information value and AoI performance is required to meet the practical demands.

Motivated by the above observations, we aim to construct an efficient joint control strategy for the UAV-GU association, semantic extraction, and the UAVs' trajectory planning to minimize the time-averaged semantic-aware
age-of-information (SAoI) of all GUs in a UAV-assisted wireless network,
which comprises a BS, multiple UAVs, and a large number of GUs.
Both the GUs and the UAVs are equipped with semantic extraction modules.
The UAVs improve wireless connectivity and extend coverage by continuously adjusting trajectories.
Specifically, the UAVs hover in the air first to determine whether to perform the semantic extraction based on the received information.
If the received information has already been extracted at the GUs, it can be directly forwarded to the BS.
Otherwise, the UAVs can carry out semantic extraction based on their processing capabilities.
Subsequently, the UAVs move to the next locations based on the GUs' AoI and information value. The joint optimization of the UAV-GU association, the UAVs' trajectories, and semantic extraction strategies is spatial-temporally coupled in different time slots, leading to a high-dimensional mixed-integer dynamic program that is difficult to solve practically.
To solve this stochastic optimization problem, we first use the Lyapunov optimization to transform it into a series of per-slot control subproblems, and then propose a hierarchical learning framework to solve the subproblem in each time slot.
%As the number of the UAVs and the GUs increases, the environment becomes high-dimensional and complex. Thus,
Specifically, we use the proximal policy optimization (PPO)~\cite{PPO-paper} algorithm to adapt the UAV-GU association.
Due to the intricate coupling between semantic extraction and UAVs' deployment, we employ alternating optimization (AO) and successive convex approximation (SCA)~\cite{2014-SCA} approaches to approximate these two subproblems.
Specifically, the main contributions of this paper are summarized as follows:
\begin{itemize}
\item \emph{UAV-assisted semantic communication}:
A flexible semantic extraction scheme is provided, which adaptively selects whether to extract at the UAVs or the GUs based on real-time data size and channel conditions.
For small-size data, the GUs directly forward the raw data to the BS via UAVs' relay transmissions.
For large-size data, direct transmission of the raw data  can lead to significant transmission delays, especially under poor channel conditions.
Therefore, the GUs can prioritize the extraction of important semantic information before uploading it to the UAVs.
The UAVs can take over the semantic extraction as edge devices when the GUs lack sufficient processing capabilities.
%The UAVs can process semantic extraction from large-size data due to their strong computational power.
The UAVs not only act as relays but also serve as computational nodes, alleviating the processing burden at the GUs.
\item \emph{Lyapunov decomposition for the SAoI minimization}:
We introduce SAoI to quantify information freshness and semantic information value.
High information value helps maintain the integrity of recovered information at the receiver, while low AoI aids in more accurate and timely decision-making.
We formulate a long-term SAoI minimization problem by jointly
optimizing the UAV-GU association, the semantic extraction, and the UAVs' trajectory planning.
%The AoI dynamics are firstly transformed into virtual queues.
Then the time-averaged SAoI minimization problem is transformed into a series of per-slot control subproblems according to the Lyapunov optimization framework.
\item \emph{Hierarchical learning for UAV-GU association, semantic extraction, and UAVs' deployment:}
We propose a Lyapunov-guided hierarchical proximal policy optimization (Lya-HiPPO) algorithm to solve the time-averaged SAoI minimization problem.
The control subproblem in each time slot involves the UAV-GU association, semantic extraction, and UAVs' deployment strategies, given the GUs' AoI.
We first optimize the UAV-GU association using the model-free PPO algorithm. This can be achieved by reshaping the UAV-GU channel conditions via the UAVs' flexible flying.
Given the UAV-GU association strategy, we use the model-based optimization module to approximate the semantic extraction and the UAVs' deployment iteratively.
Simulation results demonstrate that the Lya-HiPPO algorithm not only accelerates convergence but also achieves higher rewards compared to the conventional PPO algorithm.
Moreover, the Lya-HiPPO algorithm demonstrates that semantic extraction significantly reduces AoI while maintaining the integrity of information.
\end{itemize}

Some preliminary results of this work have been reported in~\cite{2023DRL-AoI-Long}. This paper further extends the study in~\cite{2023DRL-AoI-Long} by considering the information importance at a semantic level to reduce the overall AoI.
%After semantic extraction, the small-size semantic information can be transmitted to the BS quickly, which significantly reduces the overall AoI. However, extracting small-size semantic content increases the difficulty of semantic reconstruction at the BS, lowering the information value. Low-value information may be meaningless for maintaining the integrity of information.Hence, i
In addition to analyzing the impact of semantic extraction on the overall AoI, this paper also focuses on the semantic importance.
The remainder of the paper is organized as follows:
The literature review is presented in Section II.
The system model is introduced in Section III.
The optimization problem of SAoI minimization is formulated and solved using the proposed Lya-HiPPO framework in Section IV.
The per-slot control subproblem is determined and its solution is presented in Section V.
Section VI presents the simulation results and Section VII concludes the paper.
\section{Related work}\label{sec:work}
\subsection{UAV-assisted AoI Reduction}
AoI reduction in UAV-assisted wireless networks is closely related to the UAVs' trajectories and the UAV-GU association strategies~\cite{2024-AOI-LONG}.
If the UAVs are poorly deployed, it may lead to incomplete signal coverage, which increases the data transmission delay and results in outdated information.
The proper deployment of the UAVs improves network capacity and fault tolerance.
The authors in~\cite{2022UAVsensingZhu} studied an AoI-minimal trajectory planning problem by optimizing the UAV's hovering position in a UAV-assisted IoT network.
The proposed path search algorithm can find better UAV's trajectory to keep the information fresh.
The authors in~\cite{2024-UAV-AoI-Qi} found that a double deep Q network shows outstanding ability in improving the timeliness of decision-making, reducing average AoI, and enhancing network coverage in latency-sensitive vehicular applications by effectively designing the UAV's trajectory.
The authors in~\cite{2024-UAV-AoI-Chen} aimed at minimizing the AoI and energy consumption by coordinating the UAVs' trajectories.
The DRL algorithm was developed to explore the UAVs' trajectory planning policy by utilizing sharing messages among different UAVs.
The authors in~\cite{2022-scheduling-UAV-samir} investigated the optimal online scheduling policy and dynamic UAV altitude control to maintain fresh information at the BS.
The authors proposed a DRL method to achieve the reduced expected AoI even under poor channel conditions.
The authors in~\cite{2024-UAV-scheduling-AoI-Sun} aimed to optimize  both the AoI and transmission rate in IRS-assisted UAV networks through the joint optimization of the transmission scheduling, UAV's trajectory, and IRS' phase shift matrix.
The UAV-GU association strategy generally relies on channel and AoI conditions, which are influenced by UAVs' trajectories.
Hence, their joint optimization is a key point to reduce AoI.

\subsection{UAV-assisted Semantic Communication}
In UAV-assisted communication networks, the UAVs can be rapidly deployed and repositioned to cover areas inaccessible to traditional infrastructure.
Meanwhile, semantic communication enables the UAVs to extract and understand the GUs' semantic features, facilitating smarter decision-making and precise task execution.
This can reduce data redundancy and accelerate effective information transmission.
The UAVs' trajectories typically impact the effectiveness and accuracy of semantic extraction, while the amount of semantic information can guide more efficient UAVs'  deployment.
The authors in~\cite{2022-UAV-semantic-Kang} employed a DRL algorithm to develop a lightweight model on the front-end UAV, integrating image and channel state sensing to achieve semantic block transmission.
This approach strikes a balance between transmission latency and image classification accuracy.
The authors in~\cite{2024-UAV-semantic-Yue} proposed combining semantic information with raw visual perception to encode environmental states into a unified feature space, resulting in a more efficient understanding and description of the environment.
As a result, it enables the UAVs to autonomously navigate in complex and unknown environments.
The resource allocation at a semantic level for the UAV-assisted networks is also a research hotspot.
The authors in~\cite{2023-UAV-semantic-Kang} investigated the impact of dynamic wireless fading channels on semantic transmission and used game theory to design a multi-user resource allocation scheme. This approach enhances the personalization and anti-interference performance of semantic communication and effectively improves the quality of semantic communication services.
The authors in~\cite{2023-UAV-semantic-Wang} proposed that the UAVs can provide task-oriented semantic services to secondary users (SUs) to maximize the semantic spectrum efficiency of the secondary network.
They designed a DDQN-DDPG-based resource allocation algorithm to achieve an outstanding performance gain by  optimizing the sub-channel assignment, semantic symbols assignment, UAV's transmit beamforming, and trajectory.
The joint optimization of semantic extraction and UAVs' trajectories enables the UAVs to dynamically acquire environmental information in real-time, including the GUs' locations and demands.
%The UAVs' trajectory planning improves channel conditions for semantic information transmission, facilitating a better understanding of GUs' intentions.

\subsection{Semantic-aware AoI Minimization}
Semantic-aware information freshness refers to the ability to maintain up-to-date and meaningful information in communication systems by leveraging semantic understanding.
Unlike conventional communications that focus solely on bit-level accuracy, semantic-level communication prioritizes the important content of the information being transmitted.
The authors in~\cite{2023-AoI-semantic-Chen} proposed to improve information freshness in a time division multiple access (TDMA) system by introducing a knowledge graph-assisted semantic communication (KGSC-TDMA) approach.
Leveraging a shared knowledge graph, the KGSC-TDMA encodes the meaning of update messages into triplets, thereby reducing the data size and transmission time.
The authors in~\cite{2024-aoi-semantic-Liew} introduced a semantic communication framework to provide a realistic experience and reduce AoI, leveraging the extraction and trading of scene graphs to reduce communication costs and incentivize data transmission for metaverse services.
The authors in~\cite{2023-AoI-semantic-Meng} highlighted the significance of AoI as a crucial metric in semantic communication networks, utilizing it to capture the importance of messages in the status update system.
They discovered that reliable semantic transmission can still be achieved even under unreliable channel conditions.
Fresh information is typically more timely and accurate, providing more meaningful content.
Semantic extraction tends to favor the extraction and analysis of newer information to ensure that the extracted semantic content is more practical.

\section{System model}\label{sec:model}
\begin{figure}[t]
	\centering \includegraphics[width=0.5\textwidth]{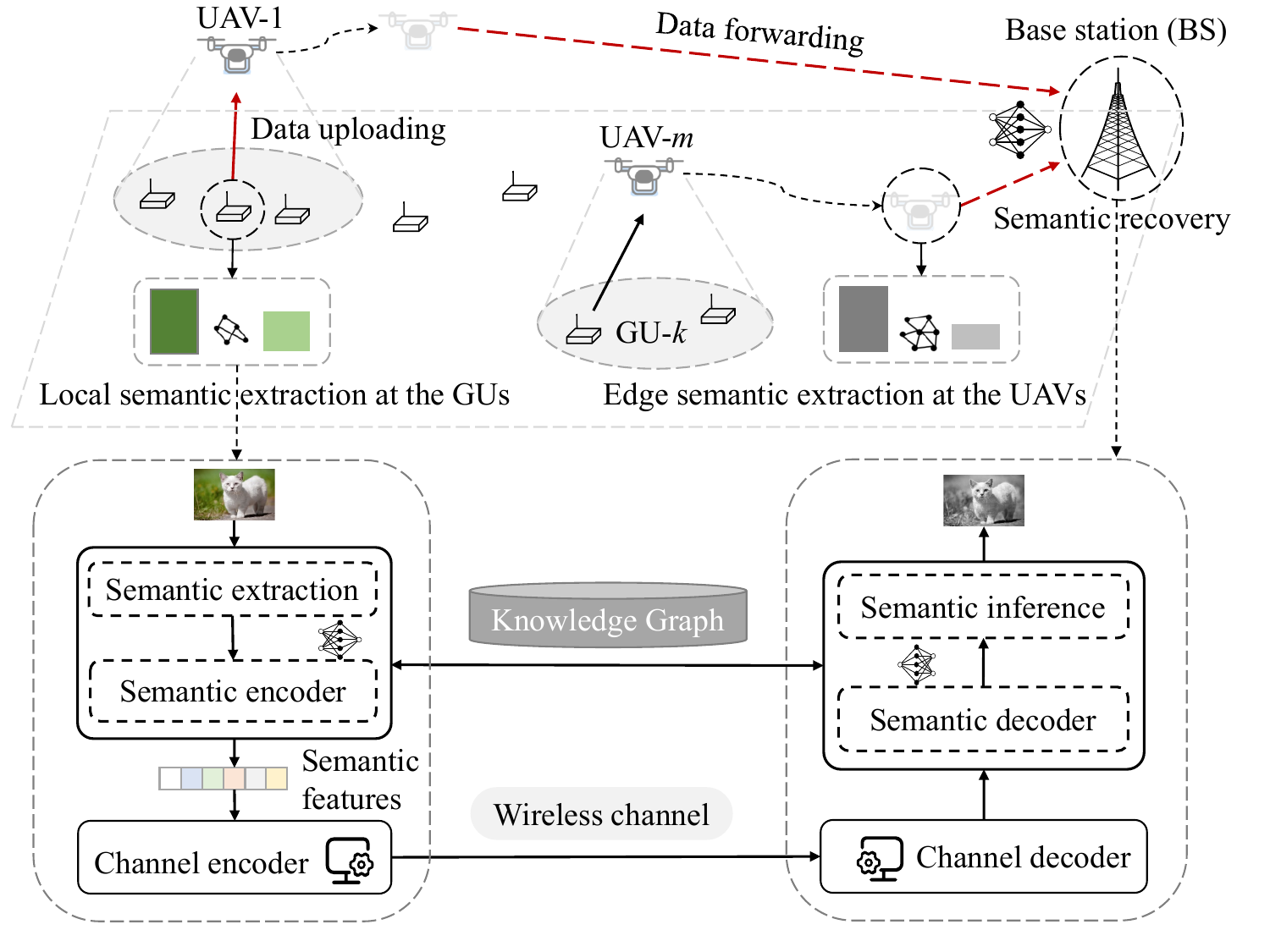}
	\caption{A UAV-assisted semantic communication network}
\label{fig_system_model}
\vspace{-2em}
\end{figure}
We consider a UAV-assisted semantic communication network with one BS, $K$ GUs and $M$ UAVs, as shown in Fig.~\ref{fig_system_model}.
The sets of the GUs and the UAVs are denoted as $\mathcal{K}\triangleq\{1,2,...,K\}$ and $\mathcal{M}\triangleq\{1,2,...,M\}$, respectively.
We consider that the GUs cannot be served by the BS directly due to the blockage of surrounding obstacles.
The UAVs are deployed to collect sensing data from $K$ GUs, randomly distributed in an open area, and then forward the collected sensing data to the BS for decision update.
%In the considered UAV-assisted semantic communication network,
Each GU randomly generates varying-size raw data.
For the small-size raw data, the UAVs can directly relay them to the BS without any processing.
 %when in superior air-to-ground channel conditions.
However, due to poor channel conditions or limited wireless resources, the transmission of large-size raw data results in high latency.
In this case, the semantic extraction facilitates shorter delay by  transmitting the most meaningful information.
Specifically, the GUs first extract the semantic information based on a local probability graph.
If the GUs lack sufficient processing capacities, the UAVs can collect large-size raw data via conventional bit-based communication and then extract the semantic information based on edge probability graphs before transmitting it to the BS.
Finally, the information recovery can be done at the BS due to its powerful processing capacity.

\subsection{Extraction-then-Transmission (ETH) Protocol}
We consider a time-slotted multi-access protocol and each frame is divided into multiple ETH time slots with equal length $t_{\max}$. The set of all time slots is denoted by $\mathcal{I}\triangleq\{1,2,...,I\}$.
We assume that all UAVs fly at the same fixed altitude and consider the 2D coordinates, where the locations of the UAV-$m$ and the GU-$k$ in the $i$-th time slot are denoted as $\bm{\ell}_m(i)=(x_m(i), y_m(i))$ and ${\bm q}_k = (x_k, y_k)$, respectively.
The location of each UAV within an ETH time slot is viewed as fixed.
We denote ${\bm q}_0 = (x_0, y_0)$ as the location of the BS's receiver antenna, which is denoted as the GU-0 for notational convenience.
Thus, the set of GUs with BS is denoted as $\tilde{\mathcal{K}} \triangleq \{\mathcal{K} \bigcup 0\}$.
As shown in Fig.~\ref{fig_system_model}, each time slot is further divided into several sub-slots $\bm t_k(i) \triangleq (t_{k,e}^l(i), t_{k,s}(i), t_{k,e}^u(i), t_{k,f}(i),t_{k,r}(i))$ for the following processes: local semantic extraction at the GUs, data uploading from the GUs to the UAVs,  edge semantic extraction at the UAVs, data forwarding from the UAVs to the BS, and semantic  recovery at the BS.
Specifically, if $t_{k,e}^l(i)=0$, it means that the GU-$k$ uploads the raw data to the UAVs.
Otherwise, the GU-$k$ extracts the semantic information from the original data in the sub-slot $t_{k,e}^l(i)$.
Based on the local semantic extraction decision, the GU-$k$ uploads either the raw data or the semantic information to the UAVs in the sub-slot $t_{k,s}(i)$.
If $t_{k,e}^u(i) = 0$, it means that the UAVs receive semantic information pre-extracted at  the GU-$k$ or raw information without further processing.
Otherwise, the UAVs act as edge devices to help extract semantic information, which reduces the burden at the GUs in the sub-slot $t_{k,e}^u(i)$.
In the sub-slot $t_{k,f}(i)$, the UAVs will forward the information from the GU-$k$ to the BS.
Finally, the BS recovers the original data from the received semantic information at the $t_{k,r}(i)$.
The information processing duration within a time slot is denoted as $\tilde{t}_k(i) = t_{k,e}^l(i) + t_{k,s}(i) + t_{k,e}^u(i) + t_{k,r}^l(i) + t_{k,r}^u(i)$.
Once the GU-$k$ is scheduled, we expect that its data processing can be completed within one time slot, which is limited as follows:
\begin{align}\label{con:time-new}
\tilde{t}_k(i) \leq t_{\max},~\forall k \in \mathcal{K}.
\end{align}

The UAVs' trajectories need to avoid collision and be subject to the speed limit $v_{\max}$ in each time slot $i \in \mathcal{I}$, i.e.,
\begin{subequations}\label{con:trajectory}
\begin{align}
&\!\|\bm{\ell}_m(i)\!-\!\bm{\ell}_{m'}(i)\| \!\geq\! d_{\rm{min}}, \quad \forall m, m' \in \mathcal{M} \text{ and } m\neq m',  \label{con:collision}\\
&\!\|\bm{\ell}_m(i)\!-\!\bm{\ell}_m(i-1)\| \!\leq\! t_{\max}v_{\max},\quad \forall m \in \mathcal{M}, \label{con:speed}
\end{align}
\end{subequations}
where $d_{\rm{min}}$ is the minimum distance between two UAVs to ensure safety, and $v_{\max}$ denotes the UAVs' maximum flying speed~\cite{2018multiUAV-tra-wu}.
For any $m \in \mathcal{M}$ and $k \in \mathcal{\tilde{K}}$, let $\mathbf{h}_{m,k}(i)$ denote the complex UAV-GU channel vector:
\begin{align}\label{equ:com-channel-UG}
 &\!\!\!\!h_{m,k}(i) \! = \! \frac{\sqrt{\xi}}{d_{s,m,k}(i)} \! \!\left( \! \sqrt{\frac{g_0}{g_0 \!+ \!1}}\bar{g}_{m,k}(i) \!\!+\!\! \sqrt{\frac{1}{g_0 + \!1}}\tilde{g}_{m,k}(i) \!\right)\! ,
\end{align}
where $\xi$ represents the channel power gain at the reference distance of 1 meter. The Rician factor $g_0$ combines the LoS and the Rayleigh fading components, denoted as $\bar{g}_{m,k}$ and $\tilde{g}_{m,k}$, respectively~\cite{2019Zhang-TWC_UAV}.

We define a binary variable $\beta_{m,k}(i)$ to denote the UAV-GU association strategy.
Specifically, $\beta_{m,k}(i)=1$ if the GU-$k$ is associated with the UAV-$m$ in the $i$-th time slot, and $\beta_{m,k}(i)=0$ otherwise.
To avoid interference, in the uploading phase, each UAV can serve a GU, and each GU can be associated with at most one UAV.
Hence, the constraints to define the UAV-GU association are as follows:
\begin{subequations}\label{con:association-UG-a}
\begin{align}
\!\!\!\!&\sum_{m \in \mathcal{M}}\beta_{m,k}(i)\leq 1 \text{ and } \sum_{k \in \mathcal{K}} \beta_{m,k}(i)\leq 1, \\
&\beta_{m,k}(i)\in \{0,1\}, ~\forall m \in \mathcal{M}  \text{ and } k \in \mathcal{K}.
\end{align}
\end{subequations}
We assume that the UAVs operate at different frequencies to avoid interference~\cite{2023-uav-Li}.
This allows for more efficient data transmission and reduces the likelihood of signal overlap.
%By working at distinct frequencies, each UAV can effectively process its data without disrupting the operations of other UAVs in the network.

\subsection{Semantic Information Extraction}
%\subsubsection{Semantic extraction}
Semantic extraction aims to obtain the essence of the source data.
Hence, compared to transmitting source data, the semantic feature transmissions reduce redundant data.
The source data extraction can be done locally (at the GUs) or at the edge (at the UAVs).
Once extracted, the original source data can be processed by the semantic extraction and semantic encoder module to derive semantic features based on the common task requirements and shared knowledge~\cite{2023-Yang-semantic-UAV-JSAC}.
The GU-$k$ can extract semantic information $s^l_k(i)$ from original source data $D_k(i)$ locally.
To measure the extraction level, we define a metric called semantic extraction depth, denoted as $\rho^l_k(i)\triangleq s^l_k(i)/D_k(i)$ at the GU-$k$ (or $\rho^u_{m,k}(i) \triangleq s^u_{m,k}(i)/D_k(i)$ at the UAV-$m$).
Hence, the required computation time for extracting semantic information locally can be expressed as follows:
\begin{align}\label{equ:time-local-extract}
%t^l_{k,e}(i) =\sum_{m \in \mathcal{M}}\beta_{m,k}(i)\alpha_k^l(i) y^{l,e}_k(D_k(i), \rho^l_k(i)) / f^l_k(i),
t^l_{k,e}(i) =\sum_{m \in \mathcal{M}}\beta_{m,k}(i)y^{l,e}_k(D_k(i), \rho^l_k(i)) / f^l_k(i),
\end{align}
where $f^l_k(i)$ is the allocated computing capacity (i.e.,  the CPU cycle frequency) to the GU-$k$ and $y^{l,e}_k(D_k(i), \rho^l_k(i))$ is a function denoting the required amount of CPU cycles for local semantic extraction, which can be approximated as follows~\cite{2023-Yang-semantic-UAV-JSAC}:
\begin{align}\label{equ:computation-GU}
\!\!\!y^{l,e}_k(D_k(i), \rho^l_k(i)) = y^{l,o}_k(D_k(i)) + B_{1k}\left(1-\rho^l_k(i)\right)^{B_{2k}},
\end{align}
where the first part $y^{l,o}_k(D_k(i))$ is to construct the shared knowledge based on original data. It can be modelled as a function related to $D_k(i)$, i.e., $y^{l,o}_k(D_k(i)) = B_{0k}D_k(i)$.
The second part refers to extracting semantic information with the depth $\rho^l_k(i)$, where the constants $B_{0k}, B_{1k}, B_{2k}>0$  can be obtained by function fitting via simulations~\cite{2023-Yang-semantic-UAV-JSAC}.

Repeated extraction of semantic information can result in the loss of contextual information, destroying semantic integrity~\cite{2023-semantic-Cang}.
It is important to clarify that the semantic information cannot be extracted for the second time.
Therefore, the UAVs can only perform semantic extraction if they receive the original data.
%As re-extraction could lead to a loss of crucial information or introduce errors.
Hence, we have the following constraints:
\begin{subequations}
\begin{align}
&0\leq \rho_{k}^l(i)\leq 1, 0\leq \rho_{m,k}^u(i)\leq 1, \label{con:rhoUAV}\\
&\rho_{k}^l(i)\rho_{m,k}^u(i) = 0, ~\forall k \in \mathcal{K},~ m\in \mathcal{M}\label{con:rhotwo}.
\end{align}
\end{subequations}
Similarly, for the UAV-$m$, the required computation time for extracting semantic information $s^u_{m,k}(i)$ from original data $D_k(i)$ is expressed as follows:
\begin{align}\label{equ:extraction-time-uav}
t^{u}_{k,e}(i)
=\sum\limits_{m \in \mathcal{M}} \beta_{m,k}(i)y^{u,e}_m(D_{k}(i), \rho_{m,k}^u(i)) / f^u_{m,k}(i),
\end{align}
where $y^{u,e}_m(D_{k}(i), \rho_{m,k}^u(i))$ is the required number of CPU cycles for calculating $s^u_{m,k}(i)$ from $D_{k}(i)$ and $f^u_{m,k}(i)$ is the computing capacity of UAV-$m$.
Similar to $y^{l,e}_k(D_k(i), \rho^l_k(i))$, the UAV-$m$'s computation function can be expressed by $y^{u,e}_m(D_k(i), \rho_{m,k}^u(i)) = y^{u,o}_m(D_{k}(i)) + C_{1m}(1-\rho_{m,k}^u(i))^{C_{2m}}$,
where $y^{u,o}_m(D_{k}(i)) = C_{0m}D_k(i) $ with $C_{0m}, C_{1m}, C_{2m} > 0$.
\subsection{Information Uploading and Forwarding}
The semantic information is then transmitted over wireless channels using conventional bit-based encoding and decoding techniques after being extracted, which ensures that the semantic information is efficiently packaged and reliably delivered through existing communication infrastructure.
%The GUs upload the semantic information to the UAVs, reducing the amount of transmitted data and decreasing the information uploading time. %Conversely, the GUs that have not performed semantic extraction will upload the original data directly.
The received SNR of the GU-$k$ at the UAV-$m$ can be represented as $\gamma_{m,k}(i) = p^o_k \left|h_{m,k}(i)\right|^2/\sigma_m^2$,
where $p^o_k$ is the GU-$k$'s transmit power and $\sigma_m^2$ represents the noise power.
Hence, the required time for information uploading from the GU-$k$ to UAV-$m$ is expressed as follows:
\begin{align}\label{equ:time-ori-sensing}
t_{k,s}(i)
= \sum_{m \in \mathcal{M}}\beta_{m,k}(i)\rho_k^l(i)D_k(i)/r_{m,k}^s(i).
\end{align}
%The UAVs also serve as relays to forward information to the BS.
Similarly, the information forwarding rate from the UAV-$m$ to BS can be expressed as $r^f_{m,0}(i) = \log_2 (1+p^u_m \left|h_{m,0}^{H}(i)\right|^2/\sigma_0^2)$, where $p^u_m$ is the UAV-$m$'s transmit power.
%There are three types of information forwarded by UAVs: original source data, local semantic information extracted by GUs, and edge semantic information extracted by the UAVs.
The data forwarding time from the UAV-$m$ to the BS can be reformulated in a compact form as follows:
\begin{align}\label{equ:time-forwarding}
\!\!t_{k,f}(i) \!= \!\sum_{m \in \mathcal{M}}\beta_{m,k}(i) (\rho_k^l(i) +\rho_{m,k}^u(i)) D_{k}(i)/r^f_{m,0}(i).
\end{align}

\subsection{Semantic Recovery}
After passing through the wireless channels, the noisy semantic information is decoded by the BS. Then, the BS  applies semantic inference techniques to derive original data and actively correct errors that occur during transmission.
The shared knowledge at the GUs (or UAVs) and the BS can guarantee the consistency of recovered information.
%The semantic information needs to be recovered at the BS.
The BS recovers the original data $D_k(i)$ either from local semantic information $s^l_k(i)$ or edge semantic information $s_{m,k}^u(i)$.
Hence, according to different semantic extraction depth $\rho_{k}^l(i)$ locally or $\rho^u_{m,k}(i))$ at the edge, the number of required CPU cycles at the BS can be expressed as $y^{r}_k(\rho_{k}^l(i), \rho^u_{m,k}(i))$ as follows~\cite{2023-Yang-semantic-UAV-JSAC}:
\begin{align}\label{equ:BS-recovery-GU}
y^{r}_k(\rho_{k}^l(i), \rho^u_{m,k}(i)) = B_{3k}\left(\rho^l_k(i) + \rho^u_{m,k}(i) \right)^{-B_{4k}},
\end{align}
where $B_{3k}, B_{4k} > 0$ are constant parameters through simulations.
Hence, the required computation time for recovering the original data $D_k(i)$ can be given as follows:
\begin{align}\label{equ:recovery-time}
t_{k,r}(i)&= \sum\limits_{m \in \mathcal{M}} \beta_{m,k}(i)y^{r}_k(\rho_{k}^l(i), \rho^u_{m,k}(i)) / g^l_k(i).
\end{align}
Typically, a greater value of $\rho^l_k(i)$ (or $\rho^u_{m,k}(i)$) indicates more preserved semantic information, facilitating more precise retrieval of the original data.
When $\rho^l_k(i)=0$ (or $\rho^u_{m,k}(i)=0$), it implies that no semantic information is extracted.
When $\rho^l_k(i)=1$ (or $\rho^u_{m,k}(i)=1$), the semantic extraction is straightforward.
It means that the GUs upload raw data to the UAVs using the conventional bit-based transmission.

\section{Lyapunov decomposition for SAoI Minimization}
\subsection{AoI Dynamics and Semantic Information Value}
\begin{figure}[t]
	\centering \includegraphics[width=0.4\textwidth]{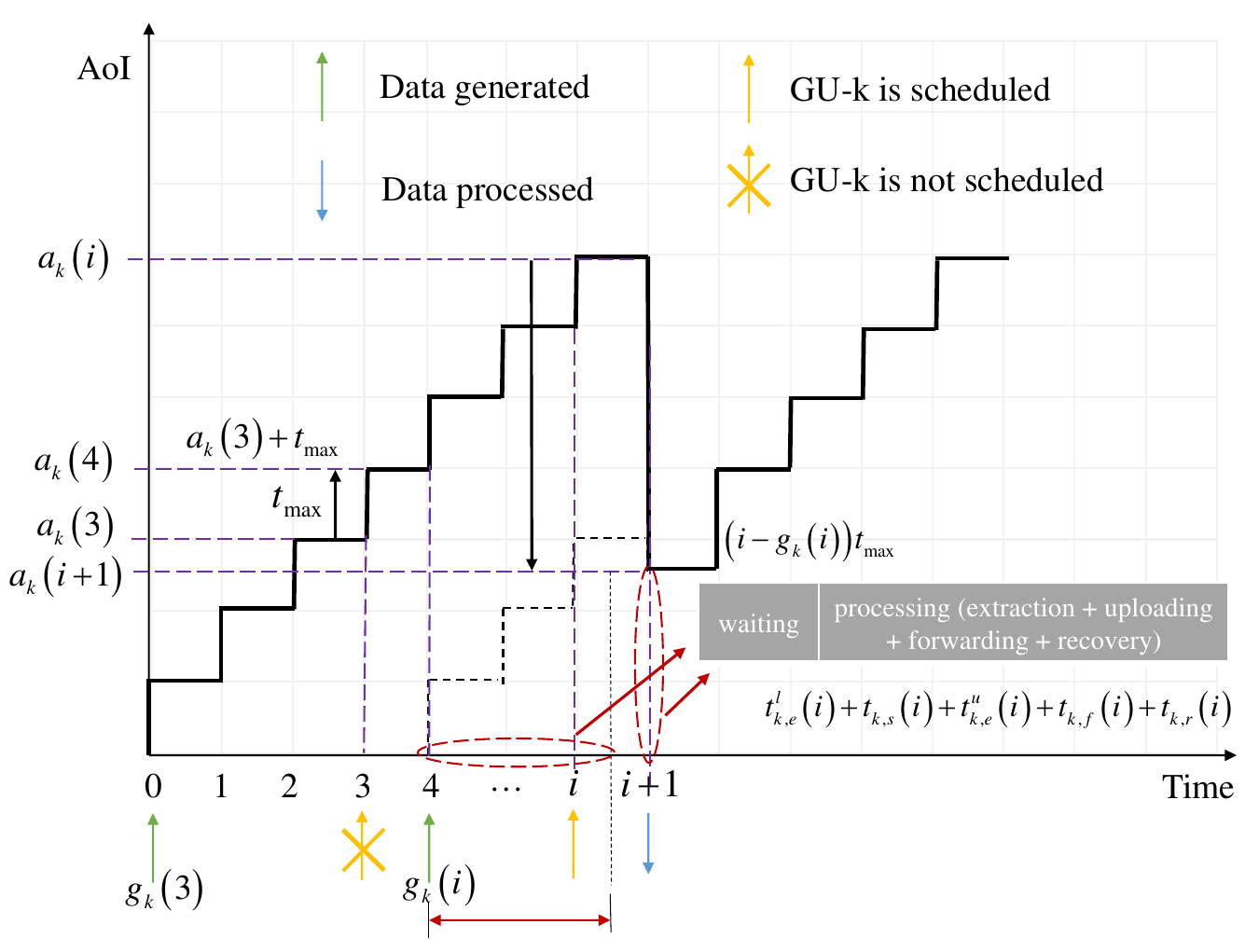}
	\caption{The GU-$k$'s AoI dynamics}\label{fig_AoI}
\vspace{-2em}
\end{figure}
The GU-$k$'s AoI is denoted by the difference between the current time and when the data for the most recently received recovery result was generated.
As shown in Fig.~\ref{fig_AoI}, if the GU-$k$ is not associated with any UAV, i.e., $\beta_{m,k}=0$, it cannot upload the data to the BS via the UAVs.
Hence, the GU-$k$'s AoI in the next time slot increases by one time slot, defined as $a_k(i+1) = a_k(i) +t_{\max}$.
If $\beta_{m,k}(i)=1$, the UAV-$m$ first collects the data from the GU-$k$ and then forwards it to the BS.
Therefore, the AoI value of GU-$k$ in the next time slot can be reduced and equal to sum of the waiting time and processing time of the latest data packet,  denoted as the second term in~\eqref{equ:aoi}.
Specifically, the waiting time is defined as the time elapsed from the generation and the departure of the latest data packet at the GU-$k$.
The time index $\tilde{g}_k(i)$ in~\eqref{equ:aoi} represents the time index of the data generated by the GU-$k$ scheduled at the $i$-th time slot.
Therefore, the total waiting time is given by $(i-\tilde{g}_k(i))t_{\max}$.
 %i.e., the unit slot length multiplied by the number of waiting time slots.
The processing time $\tilde{t}_{k}(i)$ is the duration for information extraction by the  GU-$k$ and then recovered  at the BS.
%, which includes five parts: the local semantic extraction time $t_{k,e}^l(i)$ at the GU-$k$, the GU-$k$'s information uploading time $t_{k,s}(i)$ to the UAV-$m$, the edge semantic extraction time $t_{k,e}^u(i)$ at the UAVs, the information forwarding time $t_{k,f}(i)$ to the BS and the semantic recovery time $t_{k,r}(i)$ at the BS.
In this case, the GU-$k$'s AoI value in the next time slot is given by $a_k(i+1) = (i-\tilde{g}_k(i))t_{\max} + \tilde{t}_{k}(i)$.
For notational convenience, the AoI dynamics in two cases can be rewritten in a compact form as follows:
\begin{align} \label{equ:aoi}
a_k(i+1) &= \left(1-\sum_{m \in \mathcal{M}}\beta_{m,k}(i)\right)(a_k(i) + t_{\max}) \nonumber \\
 &+ \sum_{m \in \mathcal{M}}\beta_{m,k}(i)\bigg(\big(i-\tilde{g}_k(i)\big)t_{\max}
+ \tilde{t}_{k}(i) \bigg).
\end{align}
Outdated information can lead to incorrect or wrong decisions.
To avoid such issues, it is  necessary to set a long-term average AoI limit, denoted as $a_{\max}$, to guarantee the information freshness as follows:
\begin{equation}\label{con:age}
\lim_{I \rightarrow \infty} \frac{1}{I} \sum_{i=0}^{I-1} \mathbb{E}\big[a_k(i+1)\big] \leq a_{\max}, \quad \forall k \in \mathcal{K}.
\end{equation}
The expectation is taken with respect to the potential randomness of the channel states and the UAV-GU association strategies.

The GUs or the UAVs can extract the small-size semantic information from the large-size original data to significantly reduce the overall AoI.
It means that the smaller-size semantic information (i.e., with a smaller value of  $\rho_k^l(i)$ (or $\rho_{m,k}^u(i)$) results in fresher information.
However, a smaller-size semantic extraction significantly increases the difficulty of semantic recovery at the BS, which implies a lower semantic information value.
%In some communication systems that focus on human user's needs and experiences, a decrease in semantic information value may lead to poor user's experience.
%Therefore, the optimization of semantic extraction $\rho_k^l(i) (\rho_{m,k}^u(i))$ becomes crucial.
%It is intuitive that the larger $\rho_k^l(i) (\rho_{m,k}^u(i))$, i.e., shallower semantic extraction, the easier it is to recover $D_k(i)$ from $s_k^l(i) (s_{m,k}^u(i))$.
%Hence,
We model the semantic information value function according to \cite{2023-semantic-utility-wang}, which is defined as follows:
\begin{align}\label{equ:information-value}
%v_k(i) &= \!\sum_{m \in \mathcal{M}}\!\!\beta_{m,k}(i)\bigg[\alpha_k^l(i)\left(1-e^{-B_{5k}\rho_k^l(i) D_k(i)}\right) \nonumber \\
%&+ (1-\alpha_k^l(i))\alpha_m^u(i)\!\left(\!1\!-\!e^{-C_{5m}\rho_{m,k}^u(i)\tilde{D}_{m,k}(i)}\! \right) \bigg],
v_k(i) \!\!=\! \!\sum_{m \in \mathcal{M}}\!\!\beta_{m,k}(i)\bigg(1\!-\!e^{-B_{5k}(\rho_k^l(i)+\rho_{m,k}^u(i)) D_k(i)}\bigg) ,
\end{align}
where $B_{5k} \geq 0$ is the constant parameter.
From \eqref{equ:aoi} and \eqref{equ:information-value}, an increased $\rho_k^l(i)$ (or $\rho_{m,k}^u(i)$) may diminish the information freshness, whereas a reduced $\rho_k^l(i)$ (or $\rho_{m,k}^u(i)$) compromises the semantic information value. %Hence, it is essential to strike a balance between the information freshness and semantic information value.

\subsection{SAoI Minimization and Lyapunov Decomposition}
%To ensure the information freshness and semantic importance,  we introduce a novel metric called SAoI, defined as the difference between AoI and information value.
%In this paper,
We aim to minimize the time-averaged SAoI of all GUs by jointly optimizing the UAV-GU association $\bm \beta \triangleq \{\beta_{m,k}(i)\}_{k \in \mathcal{K},m \in \mathcal{M}, i \in \mathcal{I}}$,
the semantic extraction $\bm{\rm{\rho}}\triangleq \{\rho^l_k(i), \rho^u_{m,k}(i)\}_{k \in \mathcal{K}, m \in \mathcal{M}, i \in \mathcal{I}}$,
and the UAVs' trajectories $\bm \ell \triangleq \{\ell_m(i)\}_{ m \in \mathcal{M}, i \in \mathcal{I}}$.
Considering the balance in information freshness and semantic value, we define the time-averaged SAoI as follows:
\begin{align}\label{equ:averageAoI}
\!\!\!\!\bar{A}({\bm \beta}, {{\bm \rho}, {\bm \ell}})\!\!=\!\!\lim_{I\rightarrow\infty}\!\!
\frac{1}{IK}\mathbb{E}\!\left[\sum_{i \in \mathcal{I}}\!\sum_{k \in \mathcal{K}} \omega_1a_k(i\!+\!1) \!-\! \omega_2 v_k(i)\right].
\end{align}
The expectation is taken with respect to the UAV-GU association, the semantic extraction, and the UAVs' trajectory planning in the $i$-th time slot.
It is clear that the information freshness and semantic information value have complicated couplings with the above control variables.
Till this point, we can formulate the time-averaged
SAoI minimization problem as follows:
%\begin{subequations}\label{prob:original-AoI}
\begin{align}\label{obj:original-AoI}
\min_{\bm \beta, \bm{\rho}, \bm \ell} &~\bar{A}({\bm \beta}, { {\bm \rho}, {\bm \ell}}), ~~ {\rm s.t.}~ \eqref{con:time-new}- \eqref{con:age}.
\end{align}
The problem \eqref{obj:original-AoI} is challenging to solve due to the following reasons.
Firstly, the optimization of the UAV-GU association strategy is combinatorial as it defines a discrete feasible set.
Secondly, even with the fixed UAV-GU association strategy, the UAVs' trajectories and semantic extraction are spatial-temporally coupled in a dynamic problem.
%To eliminate the temporal coupling,
We first decompose the SAoI minimization into a series of per-slot sub-problems via the Lyapunov framework~\cite{2006-Lya-queue}.
The following Proposition~\ref{prop-queue} reveals a reformulation of the time-averaged constraint in~\eqref{con:age} to simplify our problem,
%The reformulation in %Proposition~\ref{prop-queue} stems from the conclusion in~\cite{2006-Lya-queue},
which provides a generalized method to approximate a stochastic inequality constraint by using a virtual queue~\cite{2006-Lya-queue}.
\begin{proposition}\label{prop-queue}
For each GU-$k$, $k \in \mathcal{K}$, we can establish a virtual queue $Q_k(i)$ with initial zero state, i.e., $Q_k(0)=0$, and the queue dynamics given by:
\begin{equation}
Q_k(i+1) =\max\big[Q_k(i)-a_{\max},0\big] + a_k(i+1).\label{equ:AoI_queue}
\end{equation}
If $Q_k(i)$ is mean rate stable, i.e., $\lim_{i\rightarrow\infty}\frac{E[|Q_k(i)|]}{i}=0$, we can ensure the satisfaction of the inequality in \eqref{con:age}.
\end{proposition}
The detailed proof of Proposition \ref{prop-queue} can be referred to~\cite{2024-AOI-LONG}.
Then, denote $\mathbf Q(i) = (Q_1(i),...,Q_K(i))$ as the state vector of all GUs' virtual AoI queues.
We define a quadratic Lyapunov function to characterize the satisfaction of the virtual AoI queues stability constraint~\eqref{con:age}. The formulation of the Lyapunov function is given as follows:
\begin{align}\label{equ:Lyapunov function}
&~{L}\big(\mathbf{Q}(i)\big) = \frac{1}{2} \sum_{k \in \mathcal{K}} (Q_k(i))^2, i \in \mathcal{I},
\end{align}
which is a non-negative quadratic form of the virtual AoI queue states. The constant $1/2$ in~\eqref{equ:Lyapunov function} will help ease our deduction and algorithm design in the following part.
A small value of ${L}\big(\mathbf{Q}(i)\big)$ means better satisfaction.
Otherwise the Lyapunov function becomes large if at least one GU's virtual AoI queue has a large state value and tends to be unstable.

Then, the one-slot Lyapunov drift function~\cite{2006-Lya-queue} is introduced to describe a small queue change between two successive time slots, formulated as follows:
\begin{align}\label{equ:Lyapunov-Drift}
&~{\Delta}_L\big(\mathbf{Q}(i)\big) = \mathbb{E}\big[{L}\big(\mathbf{Q}(i+1)\big)-{L}\big(\mathbf{Q}(i)\big)\big|\mathbf{Q}(i)\big].
\end{align}
To stabilize the virtual AoI queue $\mathbf{Q}(i)$, we aim to minimize the increment of the queue size, i.e., the Lyapunov drift ${\Delta}_L(\mathbf{Q}(i))$. We also need to minimize all GUs' SAoI to keep information fresh and the information value. Thus, we construct a weighted minimization target in each time slot as follows:
\begin{align}\label{equ:drift-penalty} T(\mathbf{Q}(i))&\triangleq{\Delta}_L(\mathbf{Q}(i)) \nonumber \\
&+ V\sum_{k \in \mathcal{K}}\mathbb{E}\big[\omega_1a_{k}(i+1)-\omega_2 v_k(i)\big|\mathbf{Q}(i)\big],
\end{align}
where the constant $V$ is a non-negative control parameter to balance each GU's SAoI and queue stability.
To this point, we can replace the stochastic objective in \eqref{con:age} with the new minimization target in \eqref{equ:drift-penalty} and focus on the per-slot control problem in each time slot with the known states of all virtual AoI queues.
\begin{proposition}\label{prop-bound}
The weighted minimization target in~\eqref{equ:drift-penalty} is upper bounded as follows:
\begin{align}\label{equ:Lya_penalty}
 T(\mathbf{Q}(i)) \leq B + U(\bm \beta(i), \bm{\rm{\rho}}(i), \bm \ell(i)),
\end{align}
where $B = \sum_{k \in \mathcal{K}}\left[(a_{\max}^2 + (a_k(i)+t_{\max})^2)/2-Q_k(i)a_{\max}\right]$ is a finite constant, and $U(\bm \beta(i), \bm{\rm{\rho}}(i), \bm \ell(i))=\sum_{k \in \mathcal{K}} \mathbb{E}\Big\{(Q_k(i)+ V\omega_1)\Big[\left(1-\sum_{m \in \mathcal{M}}\beta_{m,k}(i)\right)(a_k(i) + t_{\max})
+ \beta_{m,k}((i-\tilde{g}_k(i))t_{\max} + \tilde{t}_{k}(i)) \Big] - V\omega_2 v_k(i)|\mathbf{Q}(i)\Big\}$.
\end{proposition}
The derivation of Proposition \ref{prop-bound} is similar to that  in~\cite{2024-AOI-LONG}.
In \eqref{equ:Lya_penalty}, the constant and optimization terms have been classified. Specifically, the first term $B$ is a finite constant since $Q_k(i)$ and $a_k(i)$ are the known states at the beginning of each time slot.
The remaining part is related to the optimization variables, including the UAV-GU association, the UAVs' deployment, and semantic extraction variables.
Instead of minimizing~\eqref{equ:drift-penalty}, we now focus on the minimization of the upper bound in~\eqref{equ:Lya_penalty} in each time slot.
For simplicity, we drop the time index in the per-slot minimization problem.
By ignoring the constant term $B$ in~\eqref{equ:Lya_penalty},
%once we observe the deterministic queue state at the beginning of the $i$-th time slot, the expectation can be removed.
%Thus,
the problem~\eqref{equ:drift-penalty} can be approximated by the following problem:
\begin{subequations}\label{prob:AoI-lya}
\begin{align}\label{obj:AoI-lya}
\min_{\bm \beta, \bm{\rho}, \bm \ell}~~~& \sum_{k \in \mathcal{K}}\Bigg((Q_k+ V\omega_1)\Bigg(\Bigg(1-\sum_{m \in \mathcal{M}}\beta_{m,k}\Bigg)(a_k + t_{\max}) \nonumber \\
+ & \sum_{m \in \mathcal{M}}\beta_{m,k}((i-\tilde{g}_k)t_{\max}\! + \tilde{t}_{k}) \Bigg) \! - \! V\omega_2v_k \Bigg) \\
{\rm s.t.} ~~~&\eqref{con:time-new} - \eqref{equ:recovery-time},
\end{align}
\end{subequations}
%Instead of the stochastic optimization in \eqref{obj:original-AoI}, now we focus on the deterministic subproblem \eqref{prob:AoI-lya},
which becomes a mixed-integer problem and is still difficult to solve directly.
The optimization of the UAV-GU association strategy is challenging because it involves a large number of discrete variables.
Even with the fixed UAV-GU association strategy, the UAVs' deployment and semantic extraction are coupled and difficult to optimize jointly.

\begin{figure}[t]
	\centering \includegraphics[width=0.5\textwidth]{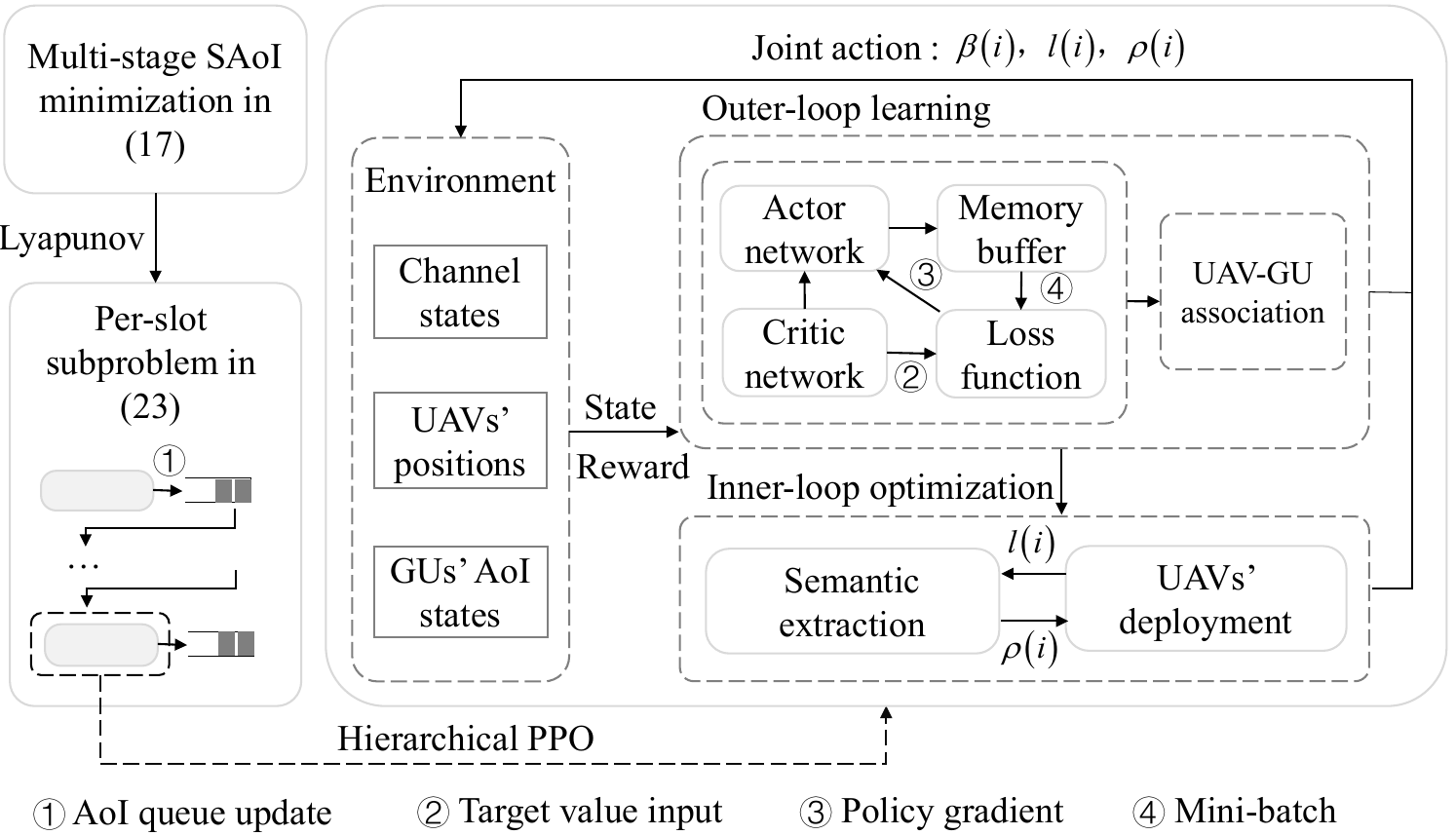}
	\caption{Lya-HiPPO algorithm framework}\label{fig_framework}
\vspace{-1.8em}
\end{figure}
\section{Per-Slot association, semantic extraction, and UAVs' deployment}
In the following, we devise a hierarchical-PPO structure framework for solving the problem~\eqref{prob:AoI-lya}, as shown in Fig. \ref{fig_framework}.
%After the Lyapunov decomposition, we further devise a hierarchical-PPO structure for it,
It mainly includes the model-free PPO for the UAV-GU association $\{\bm \beta\}$, and the model-based optimization for the UAVs' deployment and the semantic extraction $\{{\bm \ell}, {\bm \rho}\}$.
In each iteration, each UAV first determines the associated GU based on the past observations of all GUs' AoI dynamics.
Then, the model-based optimization of the UAVs' deployment and semantic extraction becomes much easier by using the SCA method.
Finally, the BS can execute the joint action $(\bm \beta, \bm{\rho}, \bm \ell)$ and update the system states.
Then, we can update the virtual AoI queue states in the next time slot.

\subsection{PPO for the UAV-GU Association}
%We design a model-free PPO algorithm for the UAV-GU association strategy.
The sets of the episode are denoted as $\mathcal{T} \triangleq\{1,2,..t.,T\}$.
The BS can prioritize the UAV-GU association strategy based on the whole system states, which includes the locations and channel states of all UAVs, as well as the AoI states of all GUs.
Once the UAV-GU association strategy is determined, the BS will distribute the decision to each UAV.
We reformulate the UAV-assisted SAoI minimization as a Markov decision process (MDP), which can be characterized by a tuple $(\mathcal{S},\mathcal{A},\mathcal{R})$.
The detailed definitions of MDP elements are given as follows:
\begin{itemize}
    \item {\textbf {State space $\mathcal{S}$}:} The set of all possible states consists of three elements: all channel gains between the UAVs and the GUs $\bm h^t \triangleq [h_{1,0}^t, ..., h_{m, K}^t]$, the UAVs' locations $\bm \ell^{t}\triangleq [\ell_{1}^t, ..., \ell_{M}^t]$, and the GUs' AoI states $\bm a^t \triangleq [a_{1}^t, ..., a_{K}^t]$.
        Therefore, at epoch $t$, the state space is represented as $\bm s^t \triangleq \{\bm h^t, \bm \ell^t,\bm a^t\}$.
        The total dimension of $\mathcal{S}$ is $MK + 2M+ K$.
    \item {\textbf {Action space $\mathcal{A}$}:} %For any DRL algorithm, the size of the action space is a major contributing factor on the algorithm's convergence and stability performance.
        Each agent chooses an action to take in the current state from its state space by using the policy $\pi: \mathcal{S}\rightarrow \mathcal{A}$.
        The action space here refers to the UAV-GU association strategy $\bm x^t \triangleq [\beta_{1,1}^t, ..., \beta_{M, K}^t]$.
        As such, the dimension of $\mathcal{A}$ is $MK$.
    \item {\textbf {Reward function $\mathcal{R}$}:} %The reward function is computed at the end of each epoch, which is the core problem of DRL in solving practical tasks.
        An effective reward function allows the system to learn a good policy.
        In this paper,
        %after the agent performs an action in the current state,
        we define $r^t(\bm s^t, \bm x^t)$ as the immediate reward in the current epoch $t$.
        %Therefore, $r^t(\bm s^t, \bm x^t)$ is a state-action function that guides the agent to select an optimal action policy.
        Since the objective is to minimize the SAoI of all GUs, we then define the immediate reward function as $r^t(\bm s^t, \bm x^t) =  U(\bm \beta^t, \bm{\rho}^t, \bm \ell^t)$.
        It also influences the DRL agent's action adaptation to minimize the long-term reward, namely, the value function as follows:
         \begin{align}
        V_{\pi}(\bm s) \triangleq \sum_{t \in \mathcal{T}}\varepsilon^t r^t(\bm s^t, \bm x^t),
        \end{align}
        where $\varepsilon \in \{0,1\}$ is the discount factor for cumulating the reward and $\pi$ denotes the policy function mapping each state $\bm s^t$ to the action $\bm x^t$.
\end{itemize}

%\subsubsection{ Proximal Policy Optimization (PPO)}
%PPO is a policy-based method that is a combination of  policy gradient (PG) and and actor-critic (AC).
Policy gradient uses the neural network with the vector $\bm \theta$ to update the policy at each round by reinforcing good behaviors.
However, finding optimal actions in large state spaces is challenging.
Thus, we use $\pi_{\bm \theta} (x^t|s^t)$ with parameters $\bm \theta$ to find the optimal parameter vector $\bm \theta$ by maximizing an average reward given by $J(\bm \theta) = E_{\bm\nu\backsim \pi_{\bm \theta}(\nu)} r^t(\bm s^t, \bm x^t)$,
%\begin{align}\label{equ:average-reward}
%J(\bm \theta) = E_{\bm\nu\backsim \pi_{\bm \theta}(\nu)} r^t(\bm s^t, \bm x^t),
%\end{align}
where $\bm\nu = \{s^1,x^1,r^1,...,s^t,x^t,r^t\}$ represents a set of states, actions, and rewards of the agent.
%Now the expected value function $J(\bm \theta)$ becomes a function of the policy parameter $\bm \theta$.
The policy gradient theorem in [35] simplifies this to $\triangledown_{\bm \theta} J(\bm \theta) = E_{\pi_{\bm \theta}}\left[\triangledown_{ \bm \theta}\log \pi_{\bm \theta}(x^t|s^t)Q^{\pi_{\bm \theta}}(x^t|s^t)\right]$,
 %$\triangledown_{\bm \theta} J(\bm \theta)$ as follows:
%\begin{align}\label{equ:gradient}
%\triangledown_{\bm \theta} J(\bm \theta) = E_{\pi_{\bm \theta}}\left[\triangledown_{ \bm \theta}\log \pi_{\bm \theta}(x^t|s^t)Q^{\pi_{\bm \theta}}(x^t|s^t)\right],
%\end{align}
where the expectation is taken over all possible state-action
pairs following the same policy $\pi_{\bm \theta}$.
%However, if we use \eqref{equ:gradient} to update $\bm \theta$, we find that it will take a long time to obtain a good policy, and the training
%data can only be used once.
However, using this method can result in slow convergence and sensitivity to step size.
%Although vanilla PG is simple to implement, it often results in a destructively huge policy update.
%Moreover, PG is more sensitive to its update step size.
%When this step size is not properly selected, its parameter update will become worse.
%As a result, the best policy is hard to be obtained.
At present, the trust region policy optimization (TRPO) has been proven to be a better way to solve above problems. In practice, TRPO is usually implemented under an actor-critic (AC) framework with a critic network (CN) parameterized by $\omega$, an old actor-network (AN) parameterized by $\bm\theta'$, and a new AN parameterized by $\bm \theta$, respectively.
TRPO addresses these issues by ensuring stable policy updates through a Kullbak-Leibler (KL)-divergence constraint. TRPO uses an advantage function $A_{\pi_{\bm \theta}}(s^t,x^t) = Q_{\pi_{\bm \theta}}(s^t,x^t) - V_{\pi_{\bm \theta}}(s^t)$ and maximizes a surrogate objective with a constraint on the KL-divergence between the old policy $\bm\theta'$ and the new policy $\bm\theta$.
While TRPO provides stability, it is computationally complex and time-consuming due to the need to solve large-size optimization problems.
Then, the policy gradient can be rewritten as $\triangledown_{\bm \theta} J(\bm \theta) = E_{\pi_{\bm \theta}}\left[\triangledown_{ \bm \theta}\log \pi_{\bm \theta}(x^t|s^t)A_{\pi_{\bm \theta}}(s^t,x^t)\right]$.
Meanwhile, TRPO uses old policy $\pi_{\bm \theta'}$ with a parameter vector $\bm \theta'$ to interact with the environment to obtain related data, including states, actions, rewards, and possible next states, and then uses these obtained data to train new policy $\pi_{\bm \theta}$ with a parameter
vector $\bm \theta$.
Therefore, we finally get the  policy gradient as follows: %$\eqref{equ:gradient_r}$:
\begin{align}\label{equ:gradient_r2}
\triangledown_{\bm \theta} J(\bm \theta)\! \!  =\! \!  E_{\pi_{\bm \theta'}}\! \! \! \left[\frac{\pi_{\bm \theta(x^t|s^t)}}{\pi_{\bm \theta'(x^t|s^t)}}\triangledown_{ \bm \theta}\log \pi_{\bm \theta}\! (x^t|s^t)A_{\pi_{\bm \theta'}}\! (s^t,x^t)\right].
\end{align}
Since the TRPO algorithm ensures monotonic policy improvement via KL divergence constraints, we optimize $\bm \theta$ by:
\begin{subequations}\label{maxprob:TRPO}
\begin{align}
\max_{\bm \theta} ~~ &E_{\pi_{\bm \theta'}}\left[\frac{\pi_{\bm \theta(x^t|s^t)}}{\pi_{\bm \theta'(x^t|s^t)}}A_{\pi_{\bm \theta'}}(s^t,x^t)\right] \\
{\rm s.t.} ~~&E_{\pi_{\bm \theta'}}[D_{KL}(\pi_{\bm \theta},\pi_{\bm \theta'})] \leq \delta_{KL},
\end{align}
\end{subequations}
where $\delta_{KL}$  is the boundary of KL divergence between $\pi_{\bm \theta'}$ and $\pi_{\bm \theta}$.
The KL divergence represents the additional cost of using an incorrect distribution $\pi_{\bm \theta}$ instead of the real distribution $\pi_{\bm \theta'}$.
PPO simplifies this by using a surrogate objective without hard constraints, with two variants: PPO-Penalty and PPO-Clip.
%PPO is faster and easier to implement than TRPO.
PPO-Penalty includes the KL divergence as a penalty term, adjusting the penalty coefficient during the training process to scale it. In this case, the objective function becomes:
\begin{align}\label{maxprob:TRPO-r}
\max_{\bm \theta} E_{\pi_{\bm \theta'}} \left[\frac{\pi_{\bm \theta(x^t|s^t)}}{\pi_{\bm \theta'(x^t|s^t)}} A_{\pi_{\bm \theta'}} - \gamma_{ \rm{KL}}D_{KL}(\pi_{\bm \theta},\pi_{\bm \theta'})\right].
\end{align}
%For PPO-Clip, to avoid the incentive for large policy update, the KL divergence and constraints are replaced by special tailoring of the objective function to reduce the difference between $\pi_{\bm \theta}$ and $\pi_{\bm \theta'}$.
%The PPO-Clip objective function is
PPO-Clip avoids large policy updates by clipping the objective function as follows:
\begin{align}\label{equ:ppo-clip}
\!\!\tilde{J}(\pi_{\bm \theta}) = E_{\pi_{\bm \theta'}}\!\left[\min(\rho_{\bm \theta}A_{\pi_{\bm \theta'}}, \!\rm{clip}(\rho_{\bm \theta}, 1 \!-\! \epsilon, 1 \!+\! \epsilon)A_{\pi_{\bm \theta'}})\right],
\end{align}
where $\rho_{\bm \theta} = \left[\frac{\pi_{\bm \theta}(x^t,s^t)}{\pi_{\bm \theta'}(x^t,s^t)} \right]$.
 The parameter $\epsilon$ is used to control the clipping range.
Let $\hat{A}(\bm \theta) = \rm{clip}(\rho_{\bm \theta}, 1 - \epsilon, 1 + \epsilon)A_{\pi_{\bm \theta'}}$, the relationship between $\hat{A}(\bm \theta)$ and $\rho_{\bm \theta}$ according to \eqref{equ:ppo-clip} is denoted as follows:
\begin{align}\label{equ:ppo-relationship}
\hat{A}(\bm \theta) =
\begin{cases}
  (1+ \epsilon)A_{\pi_{\bm \theta'}}, & \text{if } \rho_{\bm \theta} \geq 1 +  \epsilon, ~A_{\pi_{\bm \theta'}}>0,\\
  (1- \epsilon)A_{\pi_{\bm \theta'}}, & \text{if } \rho_{\bm \theta} \leq 1 -  \epsilon, ~A_{\pi_{\bm \theta'}}<0,\\
  \rho_{\bm \theta}A_{\pi_{\bm \theta'}}, & \text{otherwise}.
\end{cases}
\end{align}
When $A_{\pi_{\bm \theta'}}>0$,  it means that the current action $x^t$ is good.
So $\pi_{\bm \theta}(x^t,s^t)$ or $\rho_{\bm \theta}$ should be larger, limited by $1+ \epsilon$.
When $A_{\pi_{\bm \theta'}}<0$, it implies that the current action at cannot get a good reward. So  $\pi_{\bm \theta}(x^t,s^t)$ or $\rho_{\bm \theta}$ should be smaller, with a lower bound of $1- \epsilon$ to ensure minimal difference from small difference between $\pi_{\bm \theta}(x^t,s^t)$ and $\pi_{\bm \theta'}(x^t,s^t)$.
Therefore, when the action yields high reward, PPO restricts the over-update of $\pi_{\bm \theta}(x^t,s^t)$ to increase its probability.  Conversely, if the action obtains low reward, PPO restricts the over-update to decrease its probability.
%This limits the update range.

%The pseudocode of the proposed PPO method is summarized in \textcolor[rgb]{1.00,0.00,0.00}{Algorithm [].
%Algorithm description.}
\subsection{AO for the UAVs' Deployment and Semantic Extraction}
Given the UAV-GU association strategy  $\bm \beta \triangleq \{\beta_{m,k} \}_{k \in \mathcal{K}, m \in \mathcal{M}}$, the optimization problem~\eqref{prob:AoI-lya} can be divided into two subproblems: semantic extraction and UAVs' deployment subproblems.
%\subsubsection{Lagrangian Dual Decomposition for Semantic Extraction}
Given the UAVs' positions, and the semantic extraction range, i.e., $0 \leq \rho^l_k(i) \leq 1$ and $0 \leq \rho^u_{m,k}(i) \leq 1$, we can construct the semantic extraction subproblem.
The equality constraints $\rho_{k}^l(i)\rho_{m,k}^u(i) = 0$ make it difficult to deal with the optimization problem.
Therefore, we convert the equality constraint $\rho_{k}^l(i)\rho_{m,k}^u(i) = 0$ into a penalty term with penalty factor $\omega_0$  and incorporate it into the objective function.
Hence, the semantic extraction subproblem can be  simplified as follows:
\begin{subequations}\label{subprob:semantic}
\begin{align}
\!\!\!\!\min_{\bm \rho, \bm \varrho}&\sum_{k \in \mathcal{K}} \big[(Q_k\!+\!V\omega_1) \Gamma_1(\rho_{k}^l, \rho_{m,k}^u) \!-\!V\omega_2 \Gamma_2(\rho_{k}^l, \rho_{m,k}^u)\big] \nonumber \\
&+ \omega_0 \sum_{m \in \mathcal{M}}\sum_{k \in \mathcal{K}} \varrho_{m,k}\\
\!\!\!\!{\rm s.t.} ~&\Gamma_1(\rho_{k}^l, \rho_{m,k}^u) + \chi_{1k}\leq t_{\max}, \\
&\Phi(\rho_{k}^l, \rho_{m,k}^u)\leq \varrho_{m,k},
\end{align}
\end{subequations}
where $\Phi(\rho_{k}^l, \rho_{m,k}^u)$ is the first-order Taylor expansion of $\rho_{k}^l \rho_{m,k}^u$, denoted as follows:
\begin{align}
&\!\!\Phi(\rho_{k}^l, \rho_{m,k}^u) \!\triangleq \!\frac{1}{4} \big[(\rho_k^l \!+\!\rho_{m,k}^u )^2 \!\!-\!\big((\rho_k^l)^{(\tau)} \!\!-\!(\rho_{m,k}^u)^{(\tau)}\big)^2 \nonumber \\
&\!\!-\!\!2\big((\rho_k^l)^{(\tau)} \!\!-\!(\rho_{m,k}^u)^{(\tau)}\!\big)\big(\rho_k^l\!-\!\! \rho_{m,k}^u \!\!-\!  (\rho_k^l)^{(\tau)} \!\!+\! (\rho_{m,k}^u)^{(\tau)} \big)  \big]. \nonumber
\end{align}
The $\Gamma_1(\rho_{k}^l, \rho_{m,k}^u)$ and $\Gamma_2(\rho_{k}^l, \rho_{m,k}^u)$ are given as follows:
\begin{align}
&\!\!\Gamma_1(\rho_{k}^l, \rho_{m,k}^u) \!\!
\triangleq \!\!\sum_{m=1}^M \!\!\bigg(\psi_{1k,m} \!\!+\! (\psi_{2k,m}^s \!\!+ \! \psi_{2k,m}^f)\rho^l_k  + \psi_{2k,m}^f \rho_{m,k}^u\nonumber \\
&
+ \psi_{3k,m} + \psi_{4k,m} \big((\rho_k^l)^{-B_{4k}}+ (\rho_{m,k}^u)^{-B_{4k}}\big)(g_k^l)^{-1}\bigg), \nonumber
\end{align}
and
\begin{align}
\Gamma_2(\rho_{k}^l, \rho_{m,k}^u) \triangleq -\sum_{m =1}^M  \beta_{m,k} \big(e^{-\phi_{1k}\rho_k^l}
- e^{-\phi_{1k} \rho_{m,k}^u} \big), \nonumber
\end{align}
%\begin{align}
%\!\!\!&\Phi(\rho_{k}^l, \rho_{m,k}^u) \!\triangleq \!\frac{1}{4} \Big[(\rho_k^l \!+\!\rho_{m,k}^u )^2 \!-\!\left((\rho_k^l)^{(\tau)} \!-\!(\rho_{m,k}^u)^{(\tau)}\right)^2 \nonumber \\
%\!\!\!&-2\big((\rho_k^l)^{(\tau)} \!-\!(\rho_{m,k}^u)^{(\tau)}\!\big)\big(\rho_k^l\!-\!\! \rho_{m,k}^u \!-\!  (\rho_k^l)^{(\tau)} \!+\! (\rho_{m,k}^u)^{(\tau)} \big)  \Big], \nonumber
%\end{align}
%\begin{align}
%\!\!\!\!&\Gamma_1(\rho_{k}^l, \rho_{m,k}^u) \!\!
%\triangleq \!\!\sum_{m=1}^M \!\!\bigg[\psi_{1k,m} \!\!+\! (\psi_{2k,m}^s \!+ \! \psi_{2k,m}^f)\rho^l_k  + \psi_{2k,m}^f \!\rho_{m,k}^u\nonumber \\
%&+ \!\! \psi_{3k,m} \!+\! \psi_{5k,m} \!\left[(\rho_k^l)^{-B_{4k}} \!+ (\rho_{m,k}^u\!)^{-B_{4k}}\right](g_k^l)^{-1}\!\bigg], \nonumber
%\end{align}
%\begin{align}
%\hspace{-3em}\Gamma_2(\rho_{k}^l, \!\rho_{m,k}^u) \triangleq -\sum_{m =1}^M  \beta_{m,k} \Big[e^{-\phi_{1k}\rho_k^l}
%- e^{-\phi_{1k} \rho_{m,k}^u} \Big],\nonumber
%\end{align}
with $\psi_{1k,m} \triangleq \beta_{m,k}B_{1k}
\big(\rho^l_k\!-\!1\big)^{B_{2k}}  (f_k^l)^{-1}$,
$\psi_{2k,m}^s \triangleq \beta_{m,k}D_k (r^s_{m,k})^{-1}$, $\psi_{2k,m}^f = \beta_{m,k}D_k (r^f_{m,0})^{-1}$,
$\psi_{3k,m} \triangleq  \beta_{m,k} C_{1m}\!(\rho_{m,k}^u \!\!-\!\!1)^{C_{2m}}(f_{m,k}^u)^{-1}$, %$\psi_{4k,m} \triangleq \beta_{m,k}D_k \tilde{\alpha}_{m,k} / r^f_{m,0}$,
$\psi_{4k,m} \triangleq \beta_{m,k} B_{3k}$,
%$\psi_{6k,m} \triangleq \beta_{m,k}\tilde{\alpha}_{m,k} B_{3k}$,
$\psi_{5k,m} \triangleq \beta_{m,k} y_k^{l,o}(D_k)$,
%$\psi_{8k,m}^s \triangleq \beta_{m,k}(1-\alpha_k^l)D_k / r^s_{m,k}$,
%$\psi_{8k,m}^f \triangleq \beta_{m,k}D_k(1-\alpha_k^l) / r^f_{m,0}$,
$\psi_{6k,m} \triangleq \beta_{m,k}y_m^{u,o}(D_k)$,
$\chi_{1k} = \sum_{m \in \mathcal{M}} (\psi_{5k,m} (f_k^l)^{-1} + \psi_{6k,m}(f_{m,k}^u)^{-1})
$, $\phi_{1k} \triangleq B_{5k} D_k$.
%$\phi_{2k,m} \triangleq C_{5m}D_k $.

Since both the objective function and feasible set are convex, problem~\eqref{subprob:semantic} is a convex problem.
%Since Slater's condition is satisfied in problem~\eqref{subprob:semantic}, the strong duality holds \cite{boyd2004convex}. Thus, we can apply the dual method to obtain the Karush-Kuhn-Tucker (KKT) point.
We adopt the Lagrange dual decomposition method to find an optimal solution.
The Lagrange function of problem~\eqref{subprob:semantic} is formulated as follows:
\begin{align}\label{subprob:semantic-lagrange}
&\mathcal{L}(\bm \rho, \bm \lambda_1,\bm \lambda_2) \!=\!\! \sum_{k \in \mathcal{K}} (Q_k\!+\!V\omega_1) \Gamma_1(\rho_{k}^l, \rho_{m,k}^u) \!\!- \!\!V\omega_2 \Gamma_2(\rho_{k}^l, \rho_{m,k}^u) \nonumber\\
&+ \omega_0 \sum_{m \in \mathcal{M}}\sum_{k \in \mathcal{K}} \varrho_{m,k}
+ \sum_{k \in \mathcal{K}}\lambda_{1k}\!\left[\Gamma_1(\rho_{k}^l, \rho_{m,k}^u)+ \chi_{1k} - t_{\max}\right] \nonumber\\
&+ \sum_{m \in \mathcal{M}}\sum_{k \in \mathcal{K}}\lambda_{2k,m} \left(\Phi(\rho_{k}^l, \rho_{m,k}^u) -  \varrho_{m,k}\right).
\end{align}
Hence, the semantic extraction depth $\{\rho_k^l$, $\rho_{m,k}^u \}_{m \in \mathcal{M}, k \in \mathcal{K}}$,  penalty factor $\omega_0$, and the Lagrange multipliers $\{\bm \lambda_1, \bm \lambda_2 \}\triangleq \{\lambda_{1k}, \lambda_{2k,m}\}_{k \in \mathcal{K}, m \in \mathcal{M}}$ can be updated by the gradient descent method, where the gradients are determined as follows:
\begin{align}
\!\!&\!\!\nabla_{\rho_k^l} \!\!=\!  (Q_k \!+\!V\omega_1 \!+\! \lambda_{1k})\!\!\sum_{m \in \mathcal{M}} \!\bigg[\frac{\psi_{1k,m}B_{2k}}{\rho^l_k-1} \!+\! \psi_{2k,m}^s\!+\! \psi_{2k,m}^f\nonumber \\
&\!-\!\psi_{4k,m} B_{4k}(\rho^l_k)^{-B_{4k}-1}(g_k^l)^{-1}
 \bigg] \!\!-\!\! V\omega_2 \!\sum_{m \in \mathcal{M}}\beta_{m,k} \phi_{1k} e^{-\phi_{1k} \rho^l_k }\nonumber \\
& + \frac{\omega_0}{2}\!\sum_{m \in \mathcal{M}} \lambda_{2k,m}\big[\rho^l_k + \rho^u_{m,k}-(\rho^l_k)^{(\tau)} + (\rho^u_{m,k})^{(\tau)}\big], \label{equ:rho_gu-La}
\end{align}
\begin{align}
\vspace{-2em}
&\nabla_{\rho_{m,k}^u} =  (Q_k \!+\!V\omega_1 \!+\! \lambda_{1k})   \big[\psi_{3k,m}C_{2m}(\rho_{m,k}^u \! -\!1)^{-1}+\psi_{2k,m} \nonumber \\
&- \psi_{4k,m}B_{4k} (\rho_{m,k}^u )^{-B_{4k}-1}(g_k^
l)^{-1} \big]- \!V\omega_2 \phi_{2k,m} e^{-\phi_{1k} \rho^u_{m,k}} \nonumber  \\
& + \frac{\omega_0}{2}\!\lambda_{2k,m}\big[\rho^l_k + \rho^u_{m,k}+(\rho^l_k)^{(\tau)} - (\rho^u_{m,k})^{(\tau)}\big], \label{equ:rho_uav-La}
\end{align}
and $\nabla_{\varrho_{m,k}} = \omega_0 - \lambda_{2k,m}$, $\nabla_{\lambda_{1k}} = \Gamma_1(\rho_{k}^l, \rho_{m,k}^u)+ \chi_{1k} - t_{\max}$, $\nabla_{\lambda_{2k,m}} = \Phi(\rho_{k}^l, \rho_{m,k}^u) -  \varrho_{m,k}$.
%\begin{align}
%&\nabla_{\varrho_{m,k}} = \omega_0 - \lambda_{2k,m}, \label{equ:varho-La}\\
%&\nabla_{\lambda_{1k}} = \Gamma_1(\rho_{k}^l, \rho_{m,k}^u)+ \chi_{1k} - t_{\max}, \label{equ:lamda1-La}\\
%&\nabla_{\lambda_{2k,m}} = \Phi(\rho_{k}^l, \rho_{m,k}^u) -  \varrho_{m,k}. \label{equ:lamda2-La}
%\end{align}
Setting $\nabla_{\rho_k^l} = 0$ and $\nabla_{\rho_{m,k}^u} = 0$ in \eqref{equ:rho_gu-La} and \eqref{equ:rho_uav-La}, and we can obtain the optimal solutions $(\rho_k^l)^*$ and $(\rho_{m,k}^u)^*$.
Note that the \eqref{equ:rho_gu-La} and \eqref{equ:rho_uav-La} are monotonically increasing with respect to $\rho_k^l$ and $\rho_{m,k}^u$, respectively. Thus, the solutions $(\rho_k^l)^*$ and $(\rho_{m,k}^u)^*$ can be obtained
via the bisection method.
The penalty factor $\omega_0$ is updated by  $\omega_0^{(\tau+1)} = c \omega_0^{(\tau)}$ for some positive $c$.
The Lagrange multiplier should meet the KKT condition and we can update it by the gradient ascent method as follows:
\begin{subequations}\label{equ:lamda_update-all}
\begin{align}\label{equ:lamda_update}
\!\!&(\lambda_{1k})^{(\tau + 1)} \!\!= \!\! \min \! \big(\!\max\big(\big(\lambda_{1k}\big)^{(\tau)} \!- \! \vartheta_1 \nabla_{\lambda_{1k}},0\big),1 \big), \\
\!\!&(\lambda_{2k,m})^{(\tau \!+\! 1)} \!\!=\! \! \min\!\big(\!\max \big(\!\big(\lambda_{2k,m}\!\big)^{(\tau)} \!\! \!\!-\! \! \vartheta_2 \nabla_{\lambda_{2k,m}},0\big),1 \big),
\end{align}
\end{subequations}
%Hence, the semantic extraction depth $\{\rho_k^l$, $\rho_{m,k}^u \}_{m \in \mathcal{M}, k \in \mathcal{K}}$ and the Lagrange multipliers $\{\lambda_k\}_{k \in \mathcal{K}}$ update equations are given as follows:
%\begin{subequations}\label{equ:update}
%\begin{align}
%&(\rho_k^l)^{\tau + 1} = \min\left(\max\left(\left(\rho_k^l \right)^{\tau} - \vartheta_1 \nabla_{\rho_k^l},0 \right),1 \right), \\
%&(\rho_{m,k}^u)^{\tau + 1} \!=\! \min\left(\!\max\left(\left(\rho_{m,k}^u\right)^{\tau} \!-\! \vartheta_2 \nabla_{\rho_{m,k}^u},0\right),1 \right), \\
%&(\lambda_k)^{\tau + 1} = \min\left(\max\left(\left(\lambda_k\right)^{\tau} - \vartheta_3 \nabla_{\lambda_k},0\right),1 \right),
%\end{align}
%\end{subequations}
where the index $\tau \geq 0$ denotes the $\tau$-th iteration.
%The parameters $\vartheta_1$,  $\vartheta_2$, and $\vartheta_3$ are the positive step sizes. We repeat three variables until convergence.
The parameters $\vartheta_1$ and $\vartheta_2$ are the positive step sizes.
%We repeat three variables until convergence.
%\begin{algorithm}
%\caption{Dual Method for Semantic extraction Depth}
%\begin{algorithmic}[1]
%
%\STATE \textbf{Initialize:} Lagrange multipliers $\lambda_k = 0$, $k = 1, 2, ..., K$, $\tau = 0$,
%\WHILE{$\tau \leq \tau_{\max}$}
%    \STATE Update the Lagrangian function $\mathcal{L}(\bm \rho, \bm \lambda_1)$ according to \eqref{subprob:semantic-lagrange}
%    \STATE Compute the gradient of the Lagrangian $\nabla_{\rho_k^l}$, $\nabla_{\rho_{m,k}^u}$, and $\nabla_{\lambda_k}$ according to \eqref{equ:gradient-all}
%    \STATE Update the semantic information extraction depths $\{(\rho_k^l)^*, (\rho_{m,k}^u)^*$ and Lagrange multiplier $(\lambda_k)^{\tau + 1}$ according to \eqref{equ:lamda_update}
%\ENDWHILE
%\end{algorithmic}
%\end{algorithm}

%\subsubsection{Computation resource allocation and UAVs' position deployment}
%\subsubsection{Successive Convex Approximation (SCA) for UAVs' Position Deployment}

To further simplify, we introduce new slack variables $\bm \eta \triangleq \{\eta_{m,k}, \eta_{m,0}\}_{m \in \mathcal{ M}, k \in \mathcal{K}}$, and let $\bm \ell_m^{old} = \bm \ell_m(i-1)$.
Thus, given the semantic extraction depth $\bm \rho\triangleq\{\rho_k^l, \rho_m^u\}_{k \in \mathcal{K}, m \in \mathcal{M}}$, the UAVs' position deployment subproblem can be equivalently transformed as follows:
%\begin{subequations}\label{subprob:deployment}
%\begin{align}
%\min_{\bm \ell}~&\sum_{k=1}^K (Q_k+V\omega_1)
%\sum_{m \in \mathcal{M}}\Gamma_4(\ell_m) \\
%{\rm s.t.} ~~&\sum_{m \in \mathcal{M}}\Gamma_4(\ell_m) + \chi_{3k}\leq t_{\max}, \label{con:time-1}\\
%&\eqref{con:trajectory},
%\end{align}
%\end{subequations}
%where $\Gamma_4(\ell_m)$  is
%\begin{subequations}
%\begin{align}
%\Gamma_4(\ell_m) \triangleq \frac{\phi_{3k,m} }{\log_2 \left(1+\frac{p_k^o\xi/\sigma^2}{\|\ell_m - q_k\|^2} \right)} + \frac{\phi_{4k,m} }{\log_2 \left(1+\frac{p_m^u\xi/\sigma^2}{\|\ell_m - q_0\|^2}\right)}, \nonumber
%\end{align}
%\end{subequations}
\begin{subequations}\label{subprob:deployment-convex}
\begin{align}
\min_{\bm \ell, \bm \eta}~&\sum_{k \in \mathcal{K}}\sum_{m \in \mathcal{M}}(Q_k+V\omega_1)
\left(\frac{\phi_{3k,m}}{\eta_{m,k}} + \frac{\phi_{4k,m}}{\eta_{m,0}}\right) \\
{\rm s.t.} ~~&\sum_{m \in \mathcal{M}} \left(\frac{\phi_{3k,m}}{\eta_{m,k}} + \frac{\phi_{4k,m}}{\eta_{m,0}}\right) + \chi_{3k}\leq t_{\max}, \label{con:time-1}\\
& d_{\min} \leq \tilde{\bm \ell}_{m,m'},\\
%& \frac{1}{\lambda_{m,k}} \leq \eta_{m,k}, \\
%& \frac{1}{\lambda_{m,0}} \leq \eta_{m,0}, \\
& \eta_{m,k} \leq E_{m,k}(\bm \ell_m), \\
& \eta_{m,0} \leq E_{m,0}(\bm \ell_m), \\
%& \frac{1}{\lambda_{m,0}} \leq \eta_{m,0},\\
%& \lambda_{m,0} \leq E_{m,0}(\ell_m),\\
&\|\bm \ell_m - \bm \ell_m^{old}\| \leq t_{\max}v_{\max},
\end{align}
\end{subequations}
where
\begin{subequations}
\begin{align}
&\phi_{3k,m} \triangleq \beta_{m,k}\rho_k^l D_k, ~\phi_{4k,m} \triangleq \beta_{m,k}  ( \rho_k^l + \rho_{m,k}^u )D_k, \nonumber \\
&\chi_{3k} \triangleq \sum_{m = 1}^M\beta_{m,k} \Big\{(y_k^{l,o}(D_k) +B_{1k}(\rho_k^l-1)^{B_{2k}} )/ f_k^l \nonumber \\
&~~~~+\big(y_m^{u,o}(D_k)+C_{1m}(\rho_{m,k}^u-1)^{C_{2m}}\big) / f_{m,k}^u \nonumber \\
&~~~~+B_{3k}\left((\rho_k^l)^{-B_{4k}} + (\rho_{m,k}^u)^{-B_{4k}}\right) / g_k^l\Big\}, \nonumber \\
&\tilde{\bm \ell}_{m,m'} \triangleq - \|\bm \ell_m^{(\tau)} - \bm \ell_{m'}^{(\tau)}\|^2 + 2\left(\bm \ell_m^{(\tau)} - \bm \ell_{m'}^{(\tau)}\right)^T \times (\bm \ell_m - \bm \ell_{m'}), \nonumber \\
& E_{m,k}(\bm \ell_m) \triangleq \! \!\frac{-\log_2(e)\gamma_{m,k}^{(\tau)}}{(r_{m,k}^s)^{(\tau)}\|\bm\ell_m^{(\tau)} \!- \!\bm q_k\|^2} \big(\|\bm \ell_m \!-\! \bm q_k\|^2 \!\!-\! \!\|\bm\ell_m^{(\tau)} \!- \!\bm q_k\|^2 \big) \nonumber \\
& ~~~~~~~~~~~~+ (r_{m,k}^s)^{(\tau)} , \nonumber \\
&E_{m,0}(\bm \ell_m) \triangleq \frac{-\log_2(e)\gamma_{m,0}^{(\tau)}}{(r_{m,0}^f)^{(\tau)}\|\bm \ell_m^{(\tau)}\! - \! \bm q_0\|^2} (\|\bm \ell_m \!-\! \bm q_0\|^2 \!-\! \|\bm \ell_m^{(\tau)}\! - \! \bm q_0\|^2)\nonumber \\
&~~~~~~~~~~~~+(r_{m,0}^f)^{(\tau)}. \nonumber
\end{align}
\end{subequations}

By using the SCA method, we can approximate the
non-convex objective function and constraints with the corresponding convex approximation terms.
Thus, the UAVs' deployment subproblem has been approximated by the convex one, i.e., subproblem~\eqref{subprob:deployment-convex}, which can be solved by the existing convex optimization toolbox, such as the CVX tool.
\begin{algorithm}
\caption{AO for semantic extraction and the UAVs' deployment}
\label{alg:SCA}
\begin{algorithmic}[1]
    \STATE Initialize UAVs' trajectories $\bm \ell^{(0)}$, given the UAV-GU association $\bm \beta$. Let Lagrange multipliers $\lambda_{1k}, \lambda_{2k,m} = 0,\forall k \in \mathcal{K}, m \in \mathcal{M}$, and outer iteration $\tau' = 0$.
    \WHILE{$\tau' \leq {\tau}^{'}_{\max}$}
        \STATE Given UAVs' trajectories $\{\bm \ell^{(\tau')}\}$ and let inner iteration $\tau = 0$
        \WHILE{$\tau \leq \tau_{\max}$}
        \STATE  Update the Lagrangian function $\mathcal{L}(\bm \rho, \bm \lambda_1, \bm \lambda_2)$ according to \eqref{subprob:semantic-lagrange}
        \STATE Compute the gradients  $\nabla_{\rho_k^l}$, $\nabla_{\rho_{m,k}^u}$, $\nabla_{\varrho_{m,k}}$, $\nabla_{\lambda_{1k}}$, and $\nabla_{\lambda_{2k,m}}$
        \STATE Update the semantic extraction depth $\{(\rho_k^l)^*, (\rho_{m,k}^u)^*$ and Lagrange multiplier $(\lambda_{1k})^{\tau + 1}$, $(\lambda_{2k,m})^{\tau + 1}$ according to \eqref{equ:lamda_update-all}
    \ENDWHILE
    \STATE Obtain the optimal solution of semantic extraction $\{\bm{\rm{\rho}}^{(\tau'+1)}\}$
        \STATE Solve the UAVs' deployment subproblem \eqref{subprob:deployment-convex} given semantic extraction depth  $\{\bm{\rm{\rho}}^{(\tau'+1)}\}$, and denote the optimal solution of UAVs' positions as $\{\bm \ell^{(\tau'+1)}\}$
        \STATE Update $\tau' = \tau' + 1$
     \ENDWHILE
    %\UNTIL{The increase of the objective value is below a threshold $\epsilon > 0$.}
    \STATE Obtain the optimal solution $ \{\bm{\rm{\rho}}^{(\tau')}, \bm {\ell}^{(\tau')}\}$.
\end{algorithmic}
\end{algorithm}

The above analysis defines each step for the model-based optimization algorithm to solve the problem in~\eqref{prob:AoI-lya} following the framework in Algorithm~\ref{alg:SCA}.
It aims to minimize the overall SAoI by alternatively optimizing semantic extraction and the UAVs' deployment, respectively.
Given the hovering positions, semantic information needs to be extracted adaptively according to the processing capacities of the GUs and the UAVs.
Hence, as shown in lines 5-7, we use the Lagrange multiplier method to obtain the optimal semantic extraction depth.
%Then, given the semantic extraction, the UAVs can navigate closer to the GUs to assist in information extraction and forwarding, which accelerates information transmission and promotes information updates.
Then, based on the current semantic extraction control, we utilize the SCA method to solve the UAVs' deployment subproblem to find the new optimal UAVs' locations, as shown in line 10.
By alternately optimizing these two variables, the model-based optimization algorithm ultimately finds the optimal UAVs' positions and semantic extraction strategy, thereby ensuring efficient data transmission and reducing all GUs' SAoI.

\section{Numerical Results}
In this section, we present simulation results to verify the performance gain of the proposed Lya-HiPPO framework.
In the simulation $3$ UAVs are considered to
serve $5$ GUs.
The BS's location in meters is given by $(900, 500)$.
The GUs are randomly distributed in a rectangular area with the dimension of $1000\times1000~m^2$ in the $(x, y)$-plane.
The default parameter settings are given as follows:  $v_{\text{max}} = 30 m/s$, $d_{\text{min}} = 30m$, $p_k = p_u = 35$ dBm, and $V = 100$.
A set of baseline schemes is also devised for comparison, i.e., the Max-AoI, Max-Value, and Without extraction schemes.
The Max-AoI scheme means that each UAV is associated with the GU with the highest AoI within its coverage and then collects the sensing data from the associated GU.
The Max-Value scheme allows the GU with the highest information value to access the uplink UAV-GU channel for data uploading.
The Without extraction scheme ensures that the GUs transmit their raw data to the BS via multiple UAVs without any data extraction.
It means that the Without extraction scheme has the greatest value of the received information at the BS, which can serve as an upper bound for the information value.

\subsection{Improved Learning Efficiency and Convergence}
\begin{figure}[t]
    \centering
    \begin{subfigure}[b]{0.24\textwidth}
        \centering
        \includegraphics[width=\textwidth]{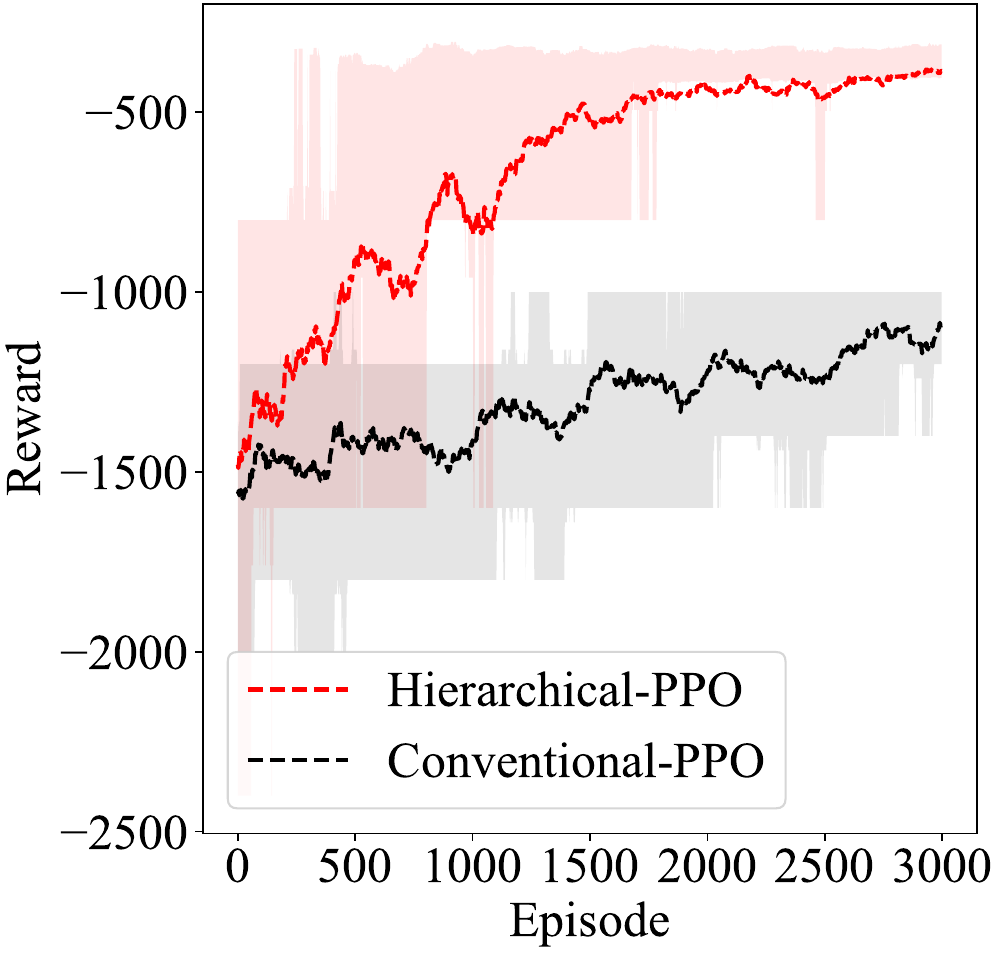}
        \caption{Model-free PPO}
        \label{Fig-convergence-single}
    \end{subfigure}
    \hfill
    \begin{subfigure}[b]{0.24\textwidth}
        \centering
        \includegraphics[width=\textwidth]{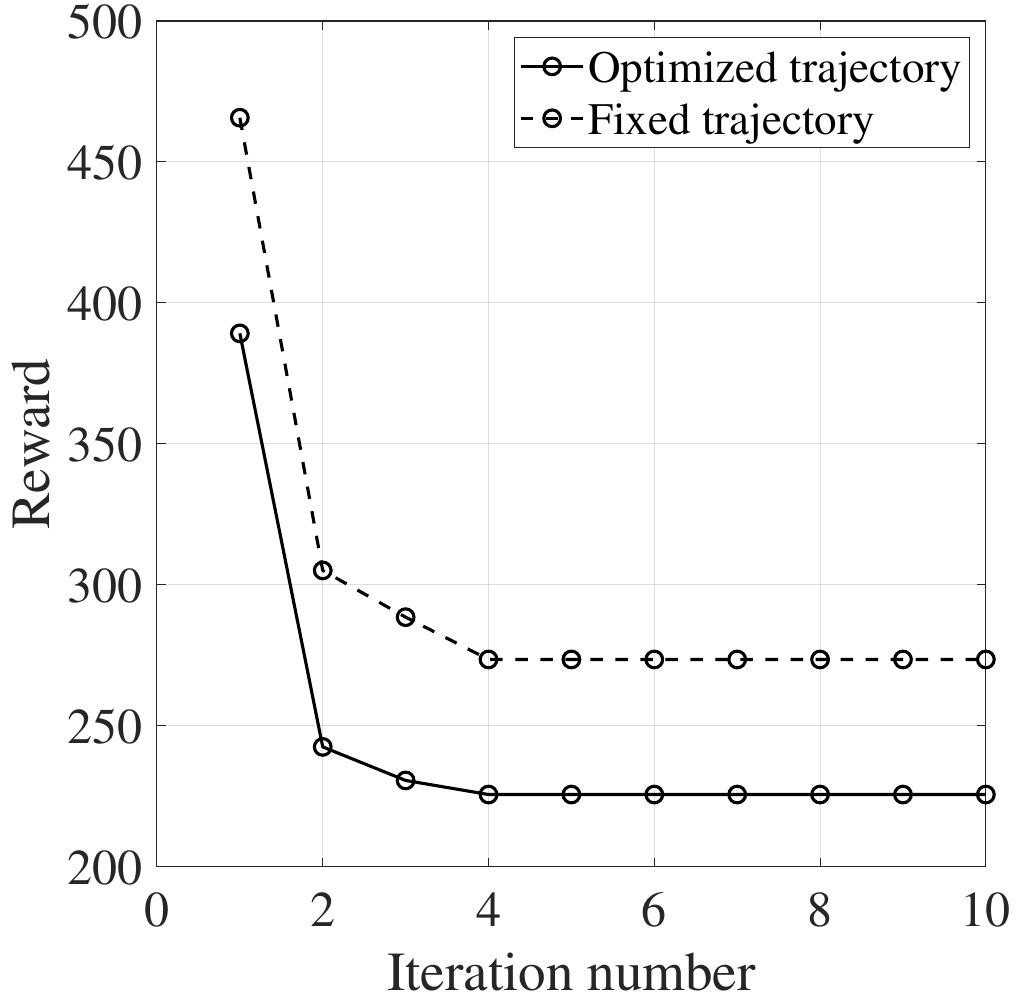}
        \caption{Model-based optimization}
        \label{Fig-convergence-opt-single}
    \end{subfigure}
    \caption{Convergence in a single time slot for model-free PPO and model-based optimization}
    \label{fig-single}
    %\vspace{-1.5em}
\end{figure}

Figures~\ref{fig-single}(a)  and \ref{fig-single}(b) show the convergence of the model-free PPO and the model-based optimization within a single time slot, respectively.
We compare the reward performance of the Hierarchical-PPO with the Conventional-PPO algorithm in Fig.~\ref{fig-single}(a), where
all decision variables $\{\beta_{m,k}, \bm \ell_m, \rho_k^l, \rho_{m,k}^u\}_{k \in \mathcal{K}, m \in \mathcal{M}}$ are adapted simultaneously in the Conventional-PPO algorithm.
The Conventional-PPO algorithm has poor convergence due to a huge action space in the mixed discrete and continuous domain.
Guided by model-based optimization, the proposed Hierarchical-PPO algorithm reduces the action space in the model-free PPO and thus achieves a higher reward performance and faster convergence.
The convergence performance of model-based optimization in a single time slot is shown in Fig.~\ref{fig-single}(b).
The reward (i.e., SAoI performance) decreases significantly within a few iterations and then converges to a stable value, which means that the complexity of the model-based  optimization algorithm is affordable.
It is observed that the model-based  optimization algorithm converges after no more than five iterations. Each
UAV first searches for a suitable position and then hovers in the air to interact with the associated GU.
Hence, the UAVs and the GUs can update their individual semantic information extraction accordingly.
Correspondingly, the reward first decreases slightly due to the change of the UAVs' positions and the semantic control, then it tends to stabilize after several iterations.

\begin{figure}[t]
    \centering
    \begin{subfigure}[b]{0.24\textwidth}
        \centering
        \includegraphics[width=\textwidth]{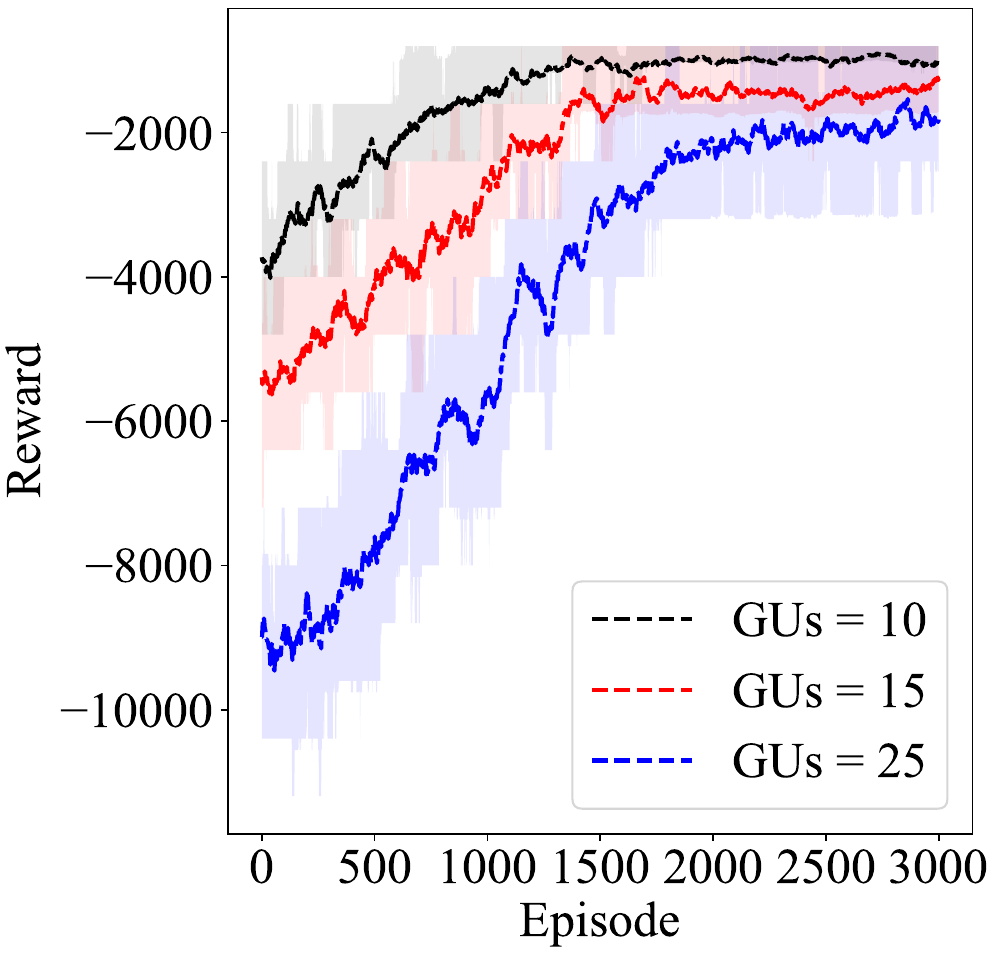}
	\caption{Single slot with varying number of GUs}
\label{Fig-convergence-gu}
    \end{subfigure}
    \hfill
    \begin{subfigure}[b]{0.24\textwidth}
        \centering
        \includegraphics[width=\textwidth]{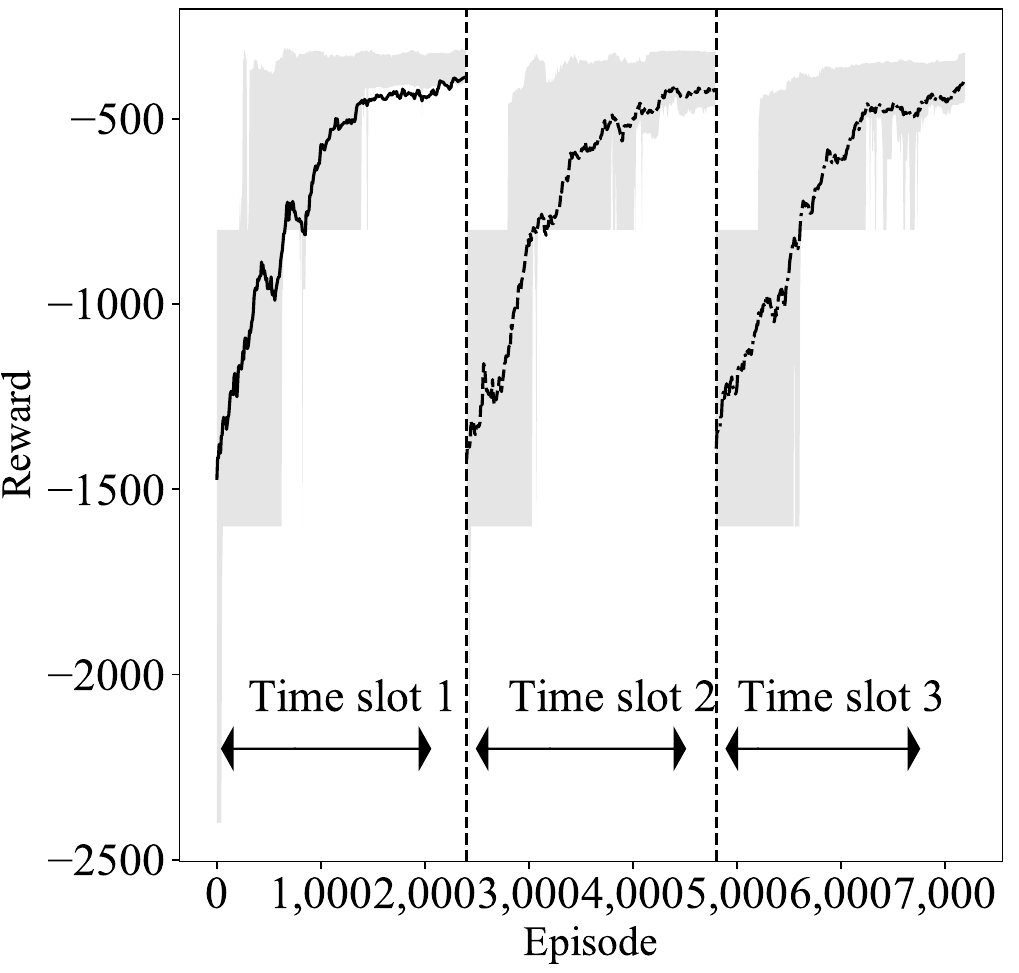}
	\caption{Multi-slot with fixed number of GUs}
\label{Fig-convergence-multi}
    \end{subfigure}
    \caption{Convergence of the Lya-HiPPO algorithm: single slot with varying number of GUs and multi-slot with fixed number of GUs}
    \label{fig-hippo}
    %\vspace{-0.5em}
\end{figure}

Figure~\ref{fig-hippo}(a) illustrates the single-slot convergence performance of the proposed Lya-HiPPO algorithm with varying number of GUs (i.e., $K=10, K = 15, K=25$).
It is observed that the convergence speed of the Lya-HiPPO algorithm varies with different numbers of GUs. Specifically, a smaller number of GUs results in faster convergence.
This is because with a smaller number of GUs, the action space is relatively small, allowing the proposed Lya-HiPPO algorithm to find the optimal action faster.
Conversely, an increase in the number of GUs leads to a larger action space, which complicates the UAV-GU association strategy.
This increased complexity requires a more extensive search and additional training episodes for the learning algorithm to converge to optimal solutions.
The multi-slot convergence of the
Lya-HiPPO framework with a fixed number of GUs (i.e., $K = 5$) is further depicted in Fig.~\ref{fig-hippo}(b).
At the beginning of each time slot, given the UAV-GU association $\{\bm \beta(i)\}$,  other actions, i.e., the UAVs' positions and the semantic extraction $\{\bm \ell(i), \bm \rho(i)\}$, can be estimated by the model-based  optimization module.
Then, the UAVs can execute the joint action, i.e., $\{\bm \beta(i), \bm \ell(i), \bm \rho(i)\}$, and the GUs' AoI can be further updated and  information value can be evaluated.
As such, the multi-slot problem can be transformed into sequential learning processes.
The updated GUs' AoIs and UAVs' positions are used as the initial states in the next time slot.
At the beginning of each time slot, the reward firstly drops significantly due to the change in the UAV-GU association strategy.
However, the reward quickly increases as the UAV-GU association strategy is adjusted by the model-free PPO algorithm.
When finding a stable joint action, the UAVs will report their locations and the semantic extraction strategy to the BS through the optimization method.
%We observe that the reward can be improved gradually at the end of each time slot, which validates the proposed Lya-HiPPO framework.

\subsection{UAVs' Trajectories and Real-time SAoI Performance}
\begin{figure}[t]
	\centering \includegraphics[width=0.4\textwidth]{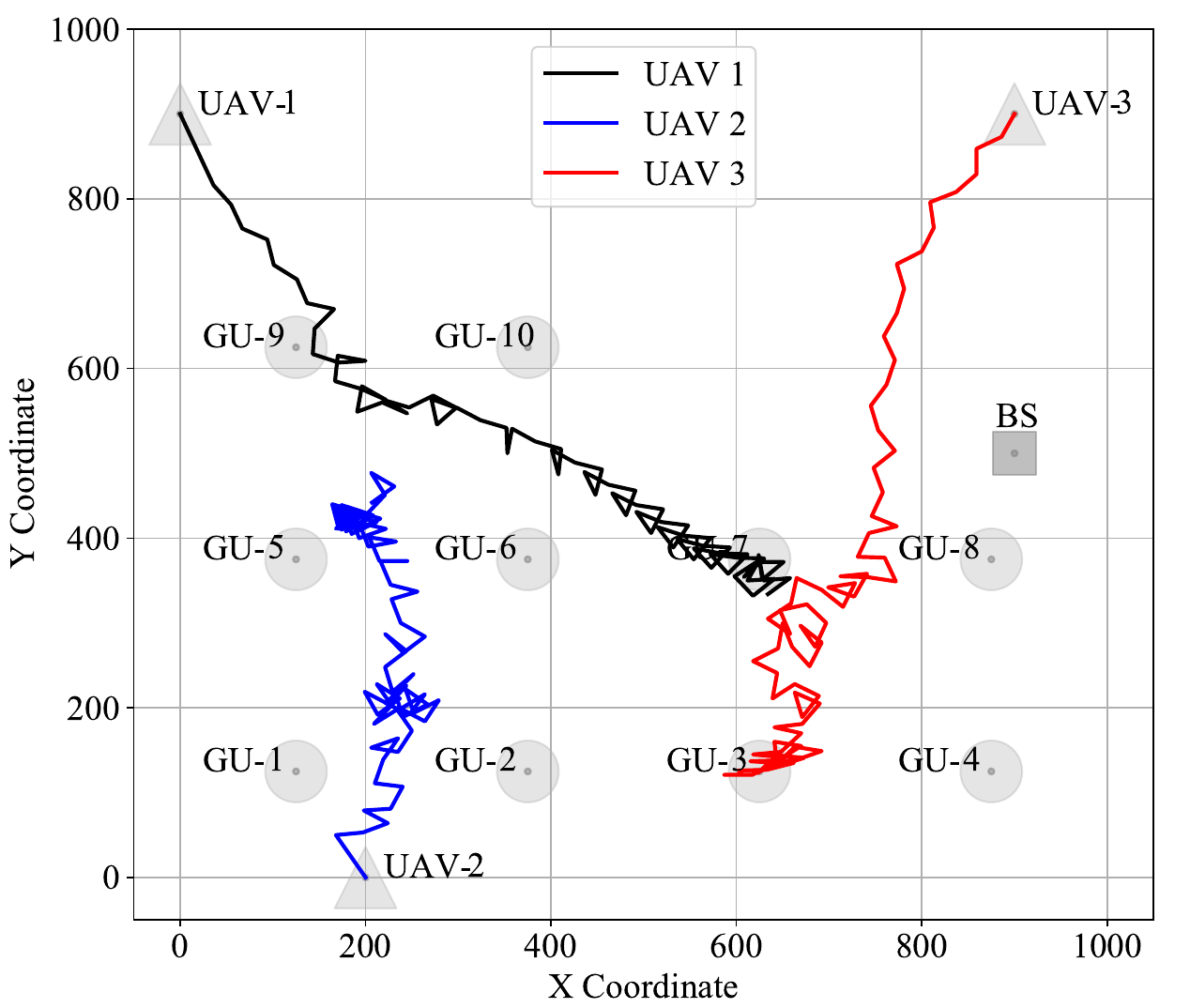}
	\caption{The UAVs' trajectories over time}\label{Fig-trajectory}
%\vspace{-1em}
\end{figure}
In Fig.~\ref{Fig-trajectory}, we demonstrate the UAVs' fast deployment and mobility along with the time.
We assume $10$ GUs is  uniformly distributed in this part.
%, and they are uniformly distributed, with equal spacing between adjacent GUs.
It is observed that each UAV can follow an optimal trajectory to serve different GUs.
The entire area is roughly divided into three parts, with each UAV servicing a specific region.
Specifically, the UAV-1 covers the GU-7, GU-9, and GU-10. The UAV-2 is responsible for the GU-1, GU-2, GU-5, and GU-6. The UAV-3 takes care of the GU-3, GU-4, and GU-8.
During the initial period of the time frame, as the UAVs are far from the GUs, they tend to accelerate toward the GUs.
Once approaching the GUs, the UAVs hover and circle nearby.
The UAVs' unsmooth trajectories are caused by the UAV-GU association strategy, which is  influenced by the GUs' AoI dynamics, spatial distribution, and information value.
Once the association is determined, each UAV flies  towards the associated GU.
However, frequent changes in these UAV-GU associations cause the UAVs to alternate among different GUs, resulting in unsmooth trajectories.

\begin{figure}[t]
    \centering
    \begin{subfigure}[b]{0.25\textwidth}
        \centering
        \includegraphics[width=\textwidth]{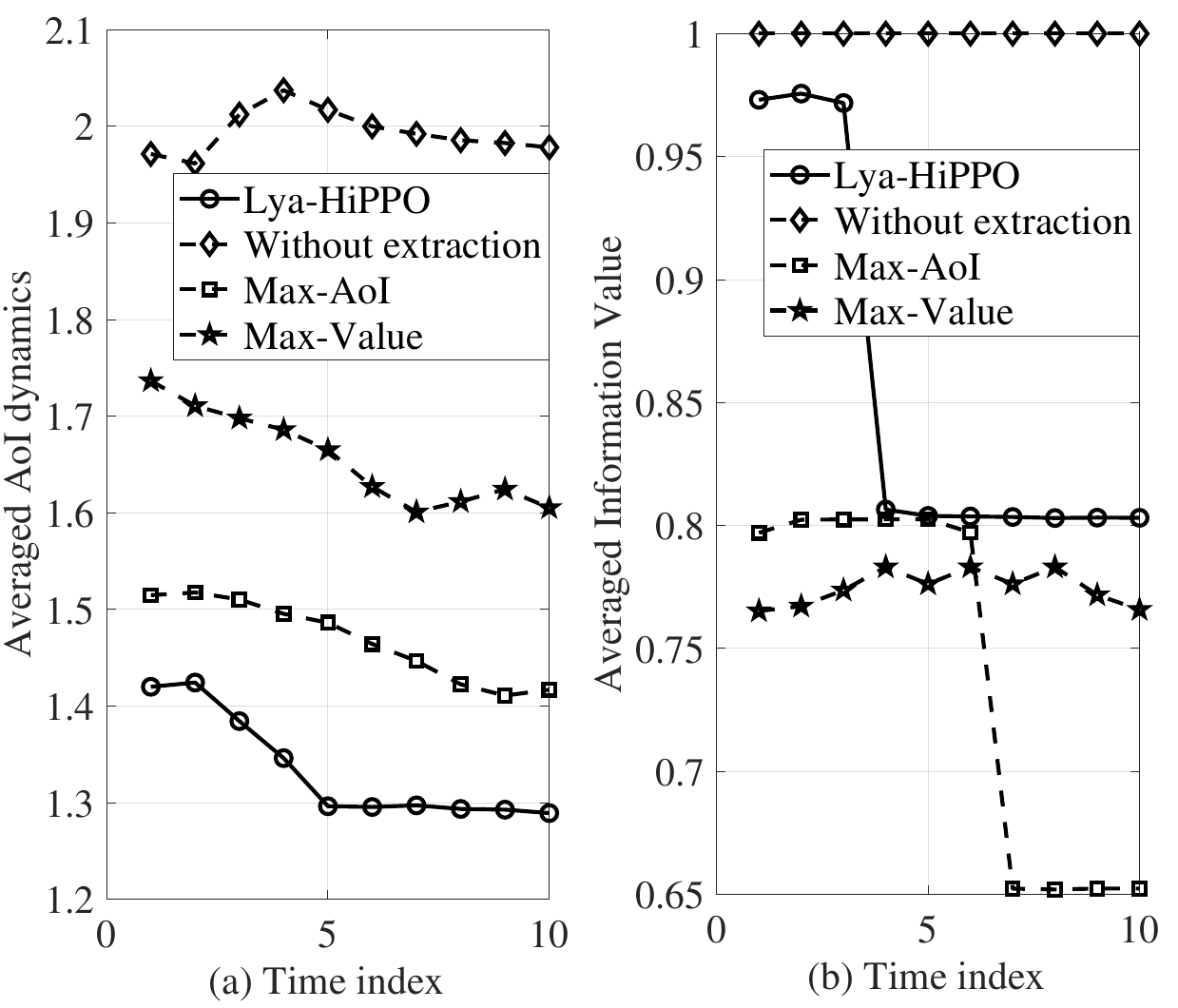}
	%\caption{AoI dynamic}
    \label{Fig-aoi-time}
    \end{subfigure}
    %\hfill
    \begin{subfigure}[b]{0.225\textwidth}
        \centering
        \includegraphics[width=\textwidth]{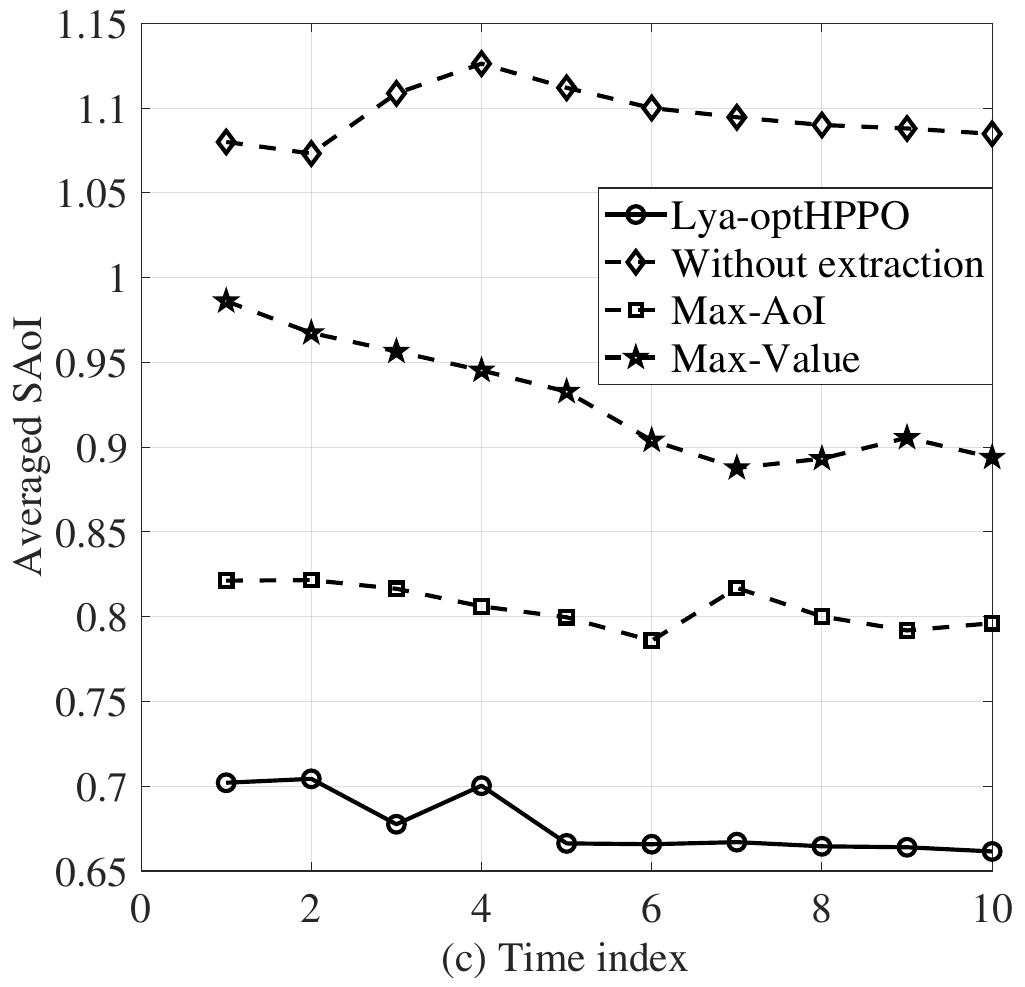}
	%\caption{Information value evolution}
    \label{Fig-value-time}
    \end{subfigure}
    \caption{Real-time SAoI with different association schemes}
    \label{fig-realtime}
    %\vspace{-2em}
\end{figure}
In Figs.~\ref{fig-realtime}(a) and \ref{fig-realtime}(b), the real-time averaged AoI dynamics and information value are plotted, respectively.
It is found that in the early stages of the time frame, the achieved overall AoI and the information value of the Lya-HiPPO scheme gradually decreases and then become stable.
The underlying reason is that in the early time slots, the UAVs are deployed away from the GUs, which results in a low GUs' uploading rate and long uploading time.
This indirectly leads to high AoI.
However, as time goes by, the UAVs gradually approach the GUs.
This behavior can improve the UAV-GU channel qualities and correspondingly shorten the uploading time, which significantly reduces the overall AoI and gradually realizes its stabilization.
Moreover, to achieve a low AoI, smaller-size semantic information is performed, leading to lower information value.
Besides, the Without extraction scheme maintains a high AoI as time goes by, which verifies that semantic extraction can improve information freshness by reducing redundant information.
The Max-AoI scheme can obtain a lower AoI than the Max-value scheme, as it effectively reduces an overall AoI by preventing excessive peak AoI. Conversely, the Max-Value scheme focuses on information value maximization, which leads to a larger-size semantic extraction.
The larger-size semantic information increases the data transmission delay, ultimately resulting in a higher AoI.
The achieved SAoI over time is illustrated in Fig.~\ref{fig-realtime}(c).
It is found that the SAoI achieved by all association schemes shows a decreasing trend, and the Lya-HiPPO scheme achieves the lowest SAoI. This further validates the previous analysis.

\subsection{SAoI Dynamics and Semantic Importance}
\begin{figure}[t]
    \centering
    \begin{subfigure}[b]{0.25\textwidth}
        \centering
        \includegraphics[width=\textwidth]{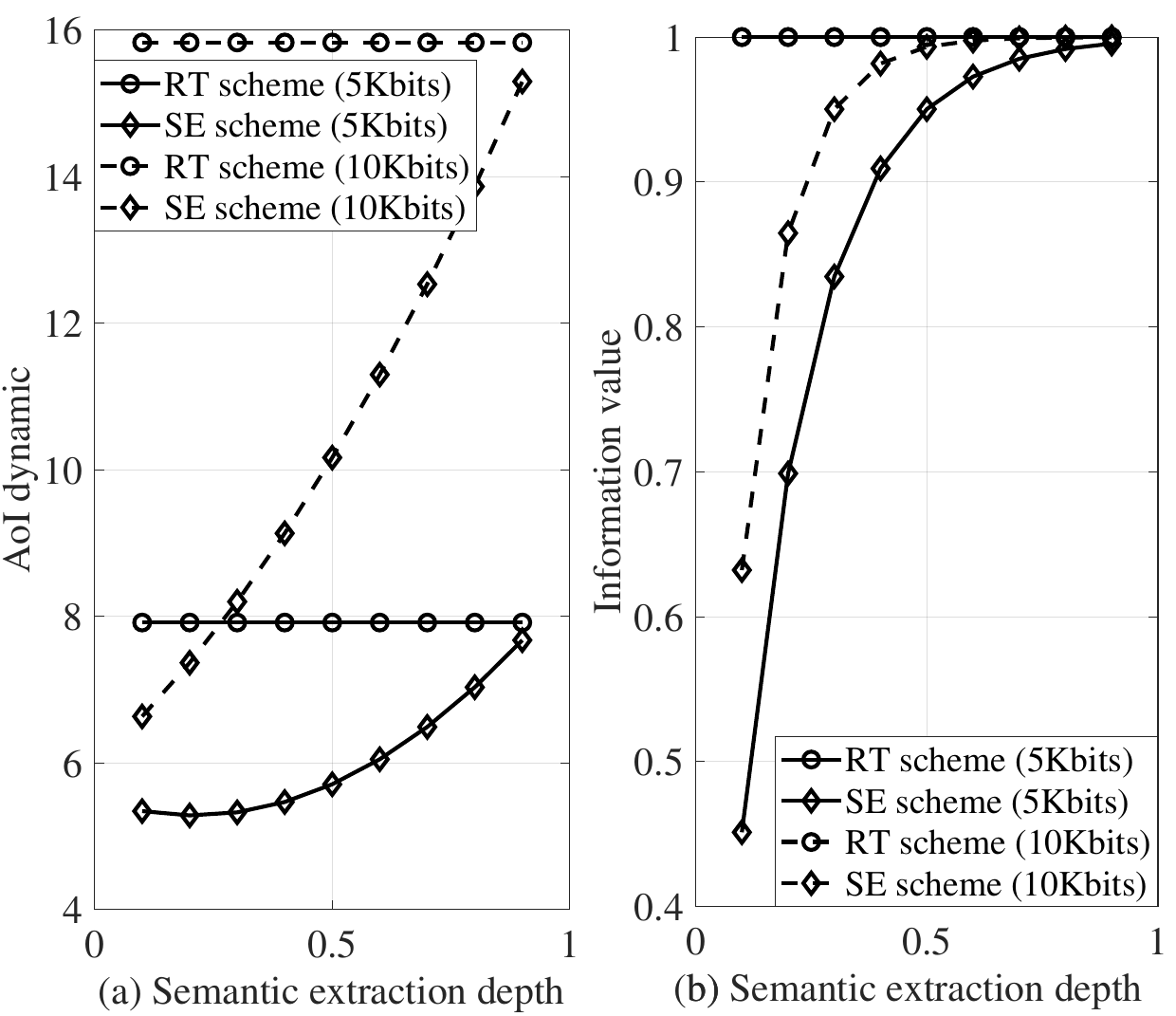}
	%\caption{AoI dynamics}
\label{Fig-AoI-depth}
    \end{subfigure}
    \hfill
    \begin{subfigure}[b]{0.215\textwidth}
        \centering
        \includegraphics[width=\textwidth]{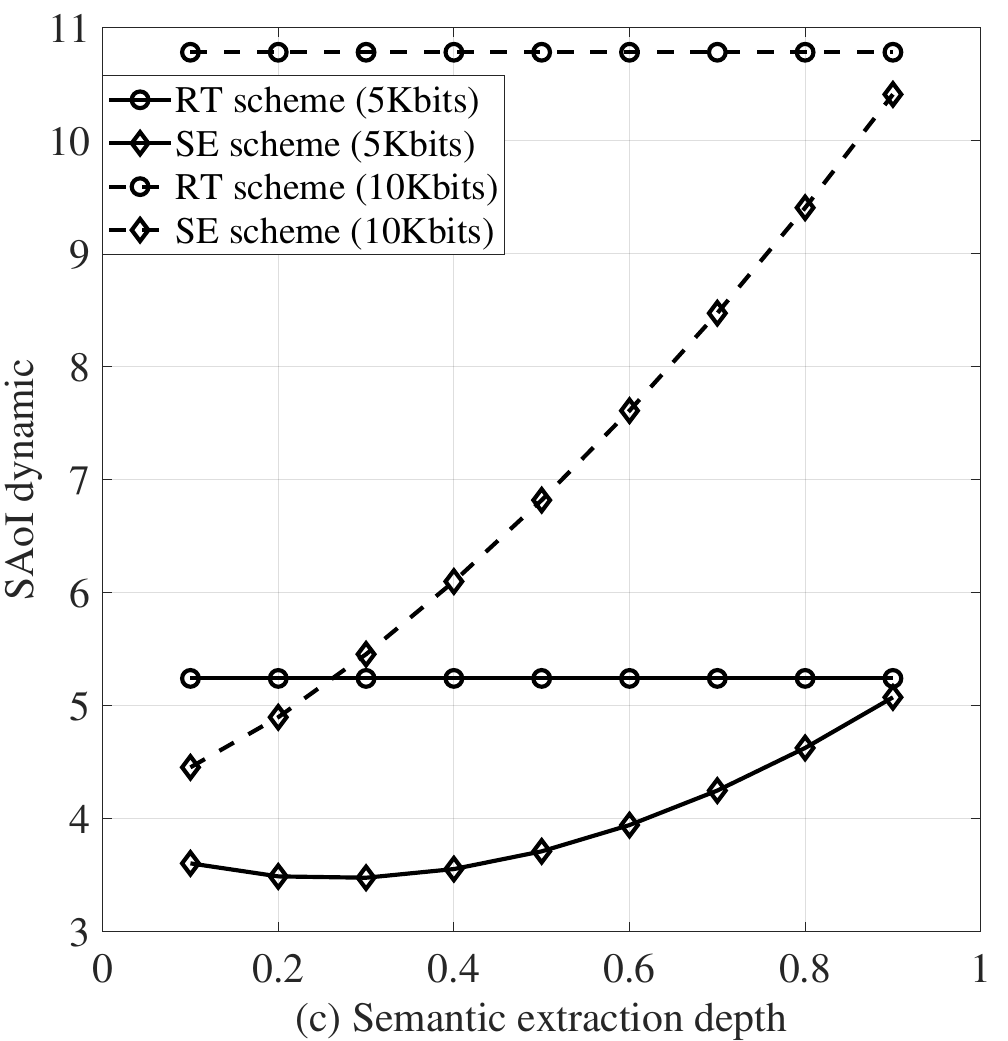}
	%\caption{Information value evolution}
\label{Fig-value-depth}
    \end{subfigure}
    \caption{Averaged SAoI with varying semantic extraction depth}
    \label{fig-depth}
    %\vspace{-1em}
\end{figure}
In Figs.~\ref{fig-depth}(a),~\ref{fig-depth}(b), and ~\ref{fig-depth}(c), we evaluate the AoI dynamics, information value, and SAoI performance with and without semantic extraction, i.e., the semantic extraction (SE) scheme and the raw data transmission (RT) scheme, given different semantic extraction depth.
Noted that a smaller value of semantic extraction depth means a deeper level of semantic extraction.
We observe that a larger semantic extraction depth leads to higher overall AoI, while the information value improves significantly.
This is because the smaller-size semantic information can reduce the transmission delay, thereby further reducing overall AoI.
However, the smaller-size semantic information increases the difficulty of semantic recovery, and the value of such information will be significantly reduced.

\begin{figure}[t]
    \centering
    \begin{subfigure}[b]{0.25\textwidth}
        \centering
        \includegraphics[width=\textwidth]{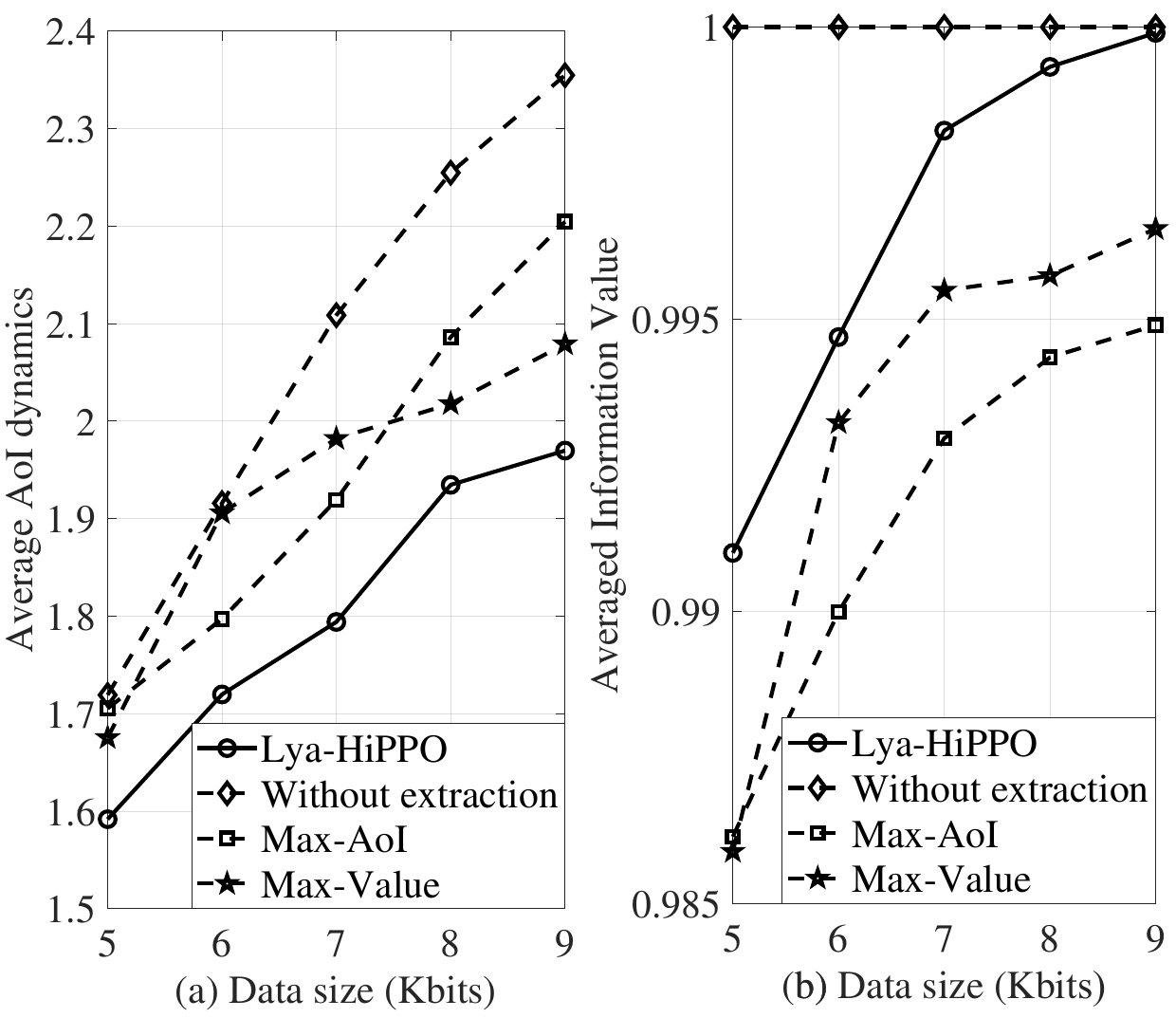}
	%\caption{Averaged AoI}
    \label{Fig-aoi-data}
    \end{subfigure}
    \hfill
    \begin{subfigure}[b]{0.21\textwidth}
        \centering
        \includegraphics[width=\textwidth]{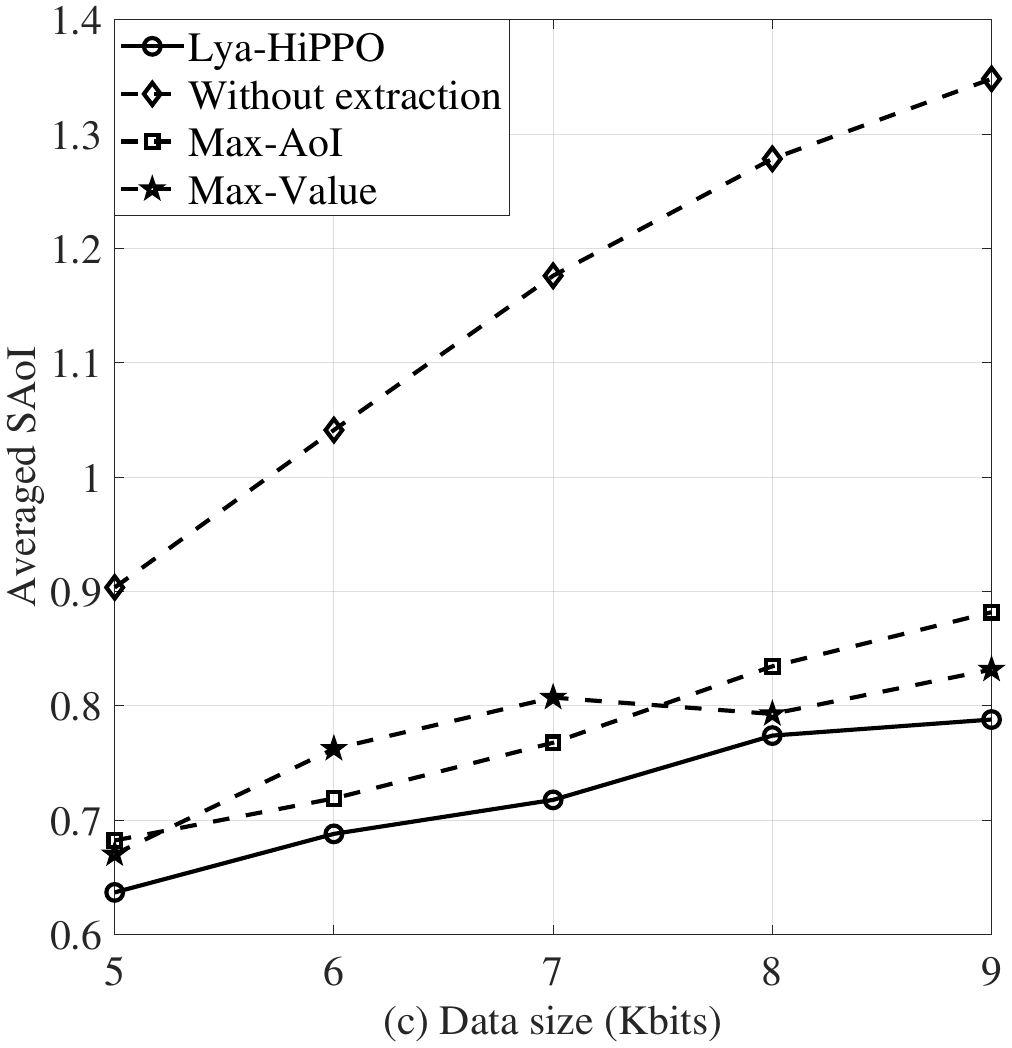}
	%\caption{Averaged information value}
    \label{Fig-value-data}
    \end{subfigure}
    \caption{Averaged SAoI with varying data size}
    \label{fig-datasize}
    %\vspace{-1.5em}
\end{figure}
Figure~\ref{fig-datasize}(a) compares the averaged AoI dynamics versus the GUs' generated data sizes.
We observe that the averaged AoI performance of the Lya-HiPPO algorithm is superior to the baselines.
This is due to the fact that transmitting more information increases both the uploading and transmission time.
A longer uploading and transmission delay leads to a higher AoI.
Besides, it is found that the Max-value scheme outperforms the Max-AoI scheme in the case of the large amount of data size.
This is because the importance of semantic information extraction becomes particularly significant when dealing with a large-size data.
As the amount of data increases, extracting semantic information requires more resources, which leads to an increased semantic extraction delay.
This will indirectly increase the overall AoI.
The Max-Value scheme enables the amount of extracted semantic information to be very close to the  original data.
This indicates that more critical information can be obtained with fewer resources, which leads to a decrease in the time required for semantic extraction, substantially reducing the overall AoI.
Fig.~\ref{fig-datasize}(b) shows the information value under varying data sizes.
The information value achieved by all schemes shows an increasing trend.
Among them, the Lya-HiPPO algorithm achieves the best information value, approaching the upper bound.
The Max-Value scheme outperforms the Max-AoI scheme, as the large amount of data can provide more comprehensive information and deeper insights to improve the information value.
By balancing overall AoI and information value, the SAoI performance is demonstrated in Fig.~\ref{fig-datasize}(c).
This further validates the superiority of the  Lya-HiPPO scheme.

\begin{figure}[t]
    \centering
    \begin{subfigure}[b]{0.25\textwidth}
        \centering
        \includegraphics[width=\textwidth]{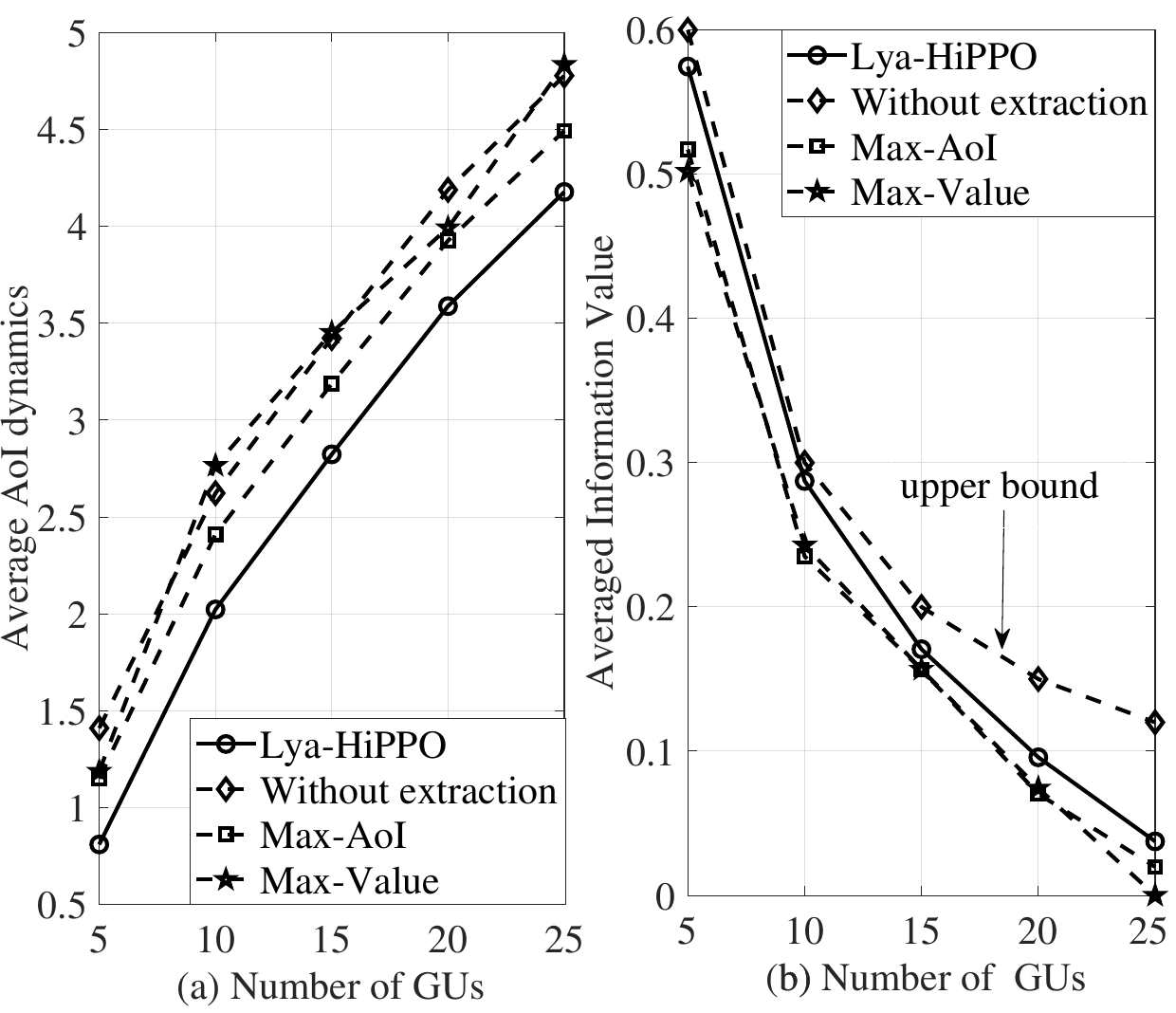}
	%\caption{Averaged AoI}
    \label{Fig-AoI-gu}
    \end{subfigure}
    \hfill
    \begin{subfigure}[b]{0.215\textwidth}
        \centering
        \includegraphics[width=\textwidth]{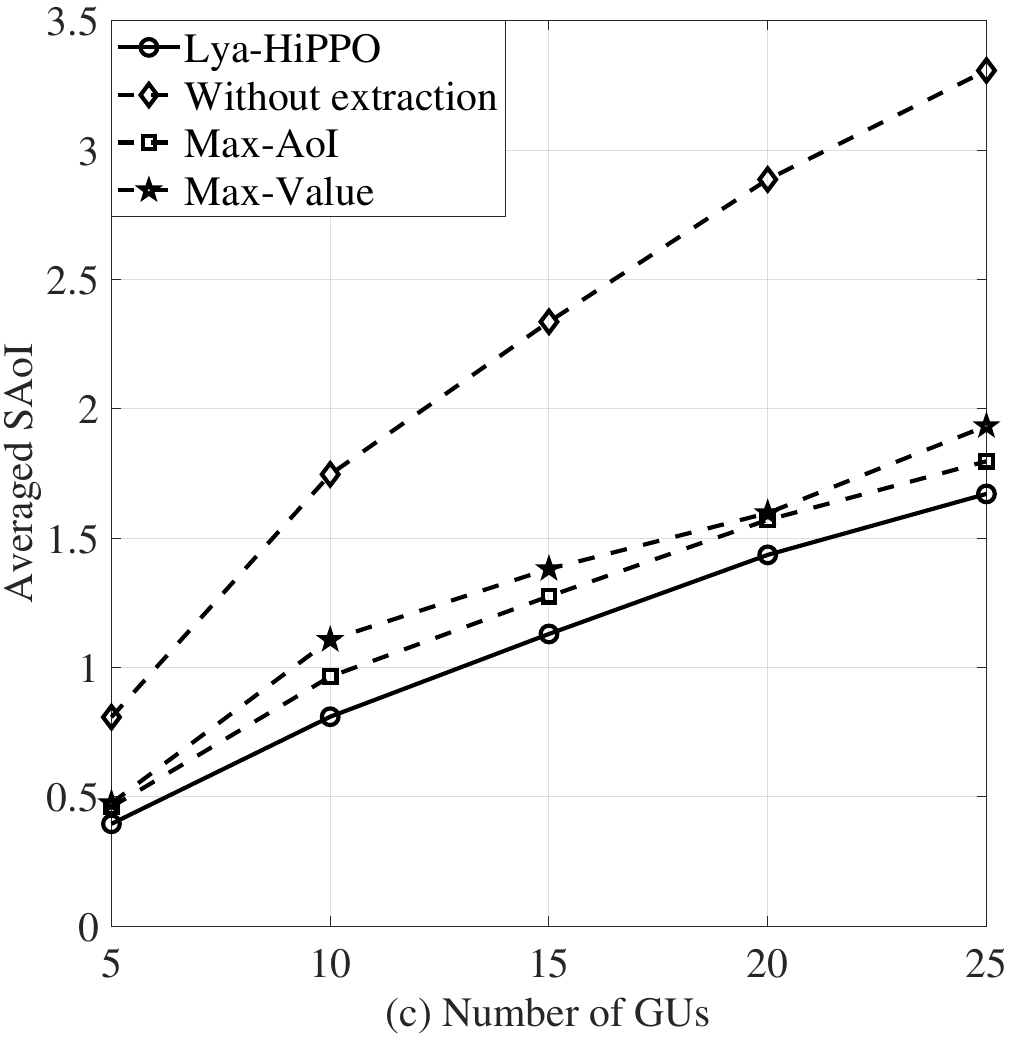}
	%\caption{Averaged information value}
    \label{Fig-value-gu}
    \end{subfigure}
    \caption{Averaged SAoI with different number of GUs}
    \label{fig-gu}
    %\vspace{-2em}
\end{figure}

We further show the GUs' AoI dynamics, the information value, and SAoI as the number of GUs increases in Figs.~\ref{fig-gu}(a), ~\ref{fig-gu}(b), and~\ref{fig-gu}(c), respectively.
It is observed that the Lya-HiPPO scheme demonstrates superior AoI and SAoI performance in Figs.~\ref{fig-gu}(a) and~\ref{fig-gu}(c).
As the number of GUs increases, the GUs' uplink contention becomes severe.
In this case, more GUs are unable to access the UAVs to upload and update data timely, which inevitably leads to higher AoI and SAoI.
In Fig.~\ref{fig-gu}(b), it is found that the averaged information value decreases as the number of GUs increases. This is because the number of UAVs is limited, the ratio of UAV-GU association decreases with the increasing number of GUs.
Meanwhile, the Lya-HiPPO scheme achieves higher information value than other baselines and closely approaches the upper bound.
By minimizing overall AoI and maximizing information value,  Fig.~\ref{fig-gu}(c) shows that the  Lya-HiPPO scheme obtains the lowest SAoI performance.
It further demonstrates the superiority of the proposed Lya-HiPPO scheme.
\section{Conclusion}
This paper has investigated the time-averaged semantic-aware age-of-information (SAoI) minimization in a semantic-aware UAV-assisted wireless network, considering the trade-off between AoI and information value.
The SAoI minimization problem has been solved by the proposed Lyapunov-guided hierarchical proximal policy optimization (Lya-HiPPO) framework.
We first used an online Lyapunov framework to decompose the long-term SAoI minimization problem into a series of per-slot control subproblems. Then, each per-slot subproblem has been solved using a hierarchical learning algorithm.
The Lya-HiPPO algorithm can keep information fresh and high information value by flexibly optimizing the UAV-GU association via the model-free PPO, then iteratively solving the semantic extraction and the UAVs' deployment via the model-based optimization approaches.
Numerical results have demonstrated that the Lya-HiPPO algorithm can efficiently leverages semantic extraction to reduce AoI while maintaining information integrity.
Our future work aims to further explore the potential of semantic communication in distributed learning, such as leveraging semantic features to avoid computationally demanding and accelerate model convergence.
\bibliographystyle{IEEEtran}
%\bibliography{your-references}
\bibliography{references}

\begin{thebibliography}{10}
\providecommand{\url}[1]{#1}
\csname url@samestyle\endcsname
\providecommand{\newblock}{\relax}
\providecommand{\bibinfo}[2]{#2}
\providecommand{\BIBentrySTDinterwordspacing}{\spaceskip=0pt\relax}
\providecommand{\BIBentryALTinterwordstretchfactor}{4}
\providecommand{\BIBentryALTinterwordspacing}{\spaceskip=\fontdimen2\font plus
\BIBentryALTinterwordstretchfactor\fontdimen3\font minus
  \fontdimen4\font\relax}
\providecommand{\BIBforeignlanguage}[2]{{%
\expandafter\ifx\csname l@#1\endcsname\relax
\typeout{** WARNING: IEEEtran.bst: No hyphenation pattern has been}%
\typeout{** loaded for the language `#1'. Using the pattern for}%
\typeout{** the default language instead.}%
\else
\language=\csname l@#1\endcsname
\fi
#2}}
\providecommand{\BIBdecl}{\relax}
\BIBdecl

\bibitem{AoI-survey}
R.~D. Yates, Y.~Sun, D.~R. Brown, S.~K. Kaul, E.~Modiano, and S.~Ulukus, ``Age
  of information: {An} introduction and survey,'' \emph{IEEE J. Sel. Areas
  Commun.}, vol.~39, no.~5, pp. 1183--1210, May 2021.

\bibitem{AoI-survey-2}
l.~Kahraman, A.~K\"{o}se, M.~Koca, and E.~Anar{\i}m, ``Age of information in
  internet of things: {A} survey,'' \emph{IEEE Internet Things J.}, vol.~11,
  no.~6, pp. 9896--9914, Oct. 2023.

\bibitem{2017Aoi-origin}
A.~Kosta, N.~Pappas, and V.~Angelakis, ``Age of information: {A} new concept,
  metric, and tool,'' \emph{Now Foundations and Trends in Netw.}, vol.~12,
  no.~3, pp. 162--259, Nov. 2017.

\bibitem{2022IRS-aoi}
Y.~Long, W.~Zhang, S.~Gong, X.~Luo, and D.~Niyato, ``{AoI}-aware scheduling and
  trajectory optimization for multi-{UAV}-assisted wireless networks,'' in
  \emph{proc. IEEE GLOBECOM}, Rio de Janeiro, Brazil, Dec. 2022, pp.
  2163--2168.

\bibitem{aimin1}
A.~Li, S.~Wu, J.~Jiao, N.~Zhang, and Q.~Zhang, ``Age of information with
  {Hybrid-ARQ}: A unified explicit result,'' \emph{IEEE Trans. Commun.},
  vol.~70, no.~12, pp. 7899--7914, Dec. 2022.

\bibitem{aimin22}
S.~Wu, Z.~Deng, A.~Li, J.~Jiao, N.~Zhang, and Q.~Zhang, ``Minimizing
  age-of-information in {HARQ-CC} aided {NOMA} systems,'' \emph{IEEE Trans.
  Wirel. Commun.}, vol.~22, no.~2, pp. 1072--1086, Feb. 2023.

\bibitem{2024Abs-UAV-real-time-Xi}
M.~Xi, H.~Dai, J.~He, W.~Li, J.~Wen, S.~Xiao, and J.~Yang, ``A lightweight
  reinforcement learning-based real-time path planning method for unmanned
  aerial vehicles,'' \emph{IEEE Internet Things J.}, vol.~11, no.~12, pp.
  21\,061--21\,071, Jan. 2024.

\bibitem{2024Abs-UAV-realtime-Liu}
R.~Liu, M.~Xie, A.~Liu, and H.~Song, ``Joint optimization risk factor and
  energy consumption in {IoT} networks with tiny{ML}-enabled internet of
  {UAVs},'' \emph{IEEE Internet Things J.}, vol.~11, no.~12, pp.
  20\,983--20\,994, Jan. 2024.

\bibitem{2024-uav-aoi-Emami}
Y.~Emami, H.~Gao, K.~Li, L.~Almeida, E.~Tovar, and Z.~Han, ``Age of information
  minimization using multi-agent {UAVs} based on {AI}-enhanced mean field
  resource allocation,'' \emph{IEEE Trans. Veh. Technol.}, pp. 1--14, Apr.
  2024, doi:10.1109/TVT.2024.3394235.

\bibitem{2024-uav-aoi-Liu}
J.~Liu, F.~Yang, X.~Wang, L.~Qu, M.~Jin, and H.~Dai, ``Joint optimization of
  charging station placement and {UAV} trajectory for fresh data collection,''
  \emph{IEEE Internet Things J.}, pp. 1--1, Apr. 2024,
  doi:10.1109/JIOT.2024.3392410.

\bibitem{2024-UAV-AoI-Pan}
J.~Pan, Y.~Li, R.~Chai, S.~Xia, and L.~Zuo, ``Age of information aware
  trajectory planning of {UAV},'' \emph{IEEE Trans. Cogn. Commun. Netw.}, pp.
  1--1, Jun. 2024, doi:10.1109/TCCN.2024.3412073.

\bibitem{2024-uav-aoi-Gong}
Z.~Gong, O.~Hashash, Y.~Wang, Q.~Cui, W.~Ni, W.~Saad, and K.~Sakaguchi,
  ``{UAV}-aided lifelong learning for {AoI} and energy optimization in
  non-stationary {IoT} networks,'' \emph{IEEE Internet Things J.}, pp. 1--1,
  May 2024, doi:10.1109/JIOT.2024.3406220.

\bibitem{2021-semantic-Xie}
H.~Xie, Z.~Qin, G.~Y. Li, and B.-H. Juang, ``Deep learning enabled semantic
  communication systems,'' \emph{IEEE Trans. Signal Process.}, vol.~69, pp.
  2663--2675, Apr. 2021.

\bibitem{2024-semantic-uav-Liew}
Z.~Q. Liew, M.~Xu, W.~Y.~B. Lim, Z.~Xiong, D.~Niyato, and D.~I. Kim,
  ``Mechanism design for semantic communication in {UAV}-assisted metaverse: A
  combinatorial auction approach,'' \emph{IEEE Trans. Veh. Technol.}, vol.~73,
  no.~2, pp. 2236--2251, Feb. 2024.

\bibitem{2023-semantic-Li}
Y.~Lil, X.~Zhou, and J.~Zhao, ``Resource allocation for semantic communication
  under physical-layer security,'' in \emph{proc. IEEE GLOBECOM}, Kuala Lumpur,
  Malaysia, Dec. 2023, pp. 2063--2068.

\bibitem{PPO-paper}
\BIBentryALTinterwordspacing
J.~Schulman, F.~Wolski, P.~Dhariwal, A.~Radford, and O.~Klimov, ``Proximal
  policy optimization algorithms,'' \emph{CoRR}, vol. abs/1707.06347, 2017.
  [Online]. Available: \url{http://arxiv.org/abs/1707.06347}
\BIBentrySTDinterwordspacing

\bibitem{2014-SCA}
\BIBentryALTinterwordspacing
M.~Razaviyayn, ``Successive convex approximation: {Analysis} and
  applications,'' Retrieved from the University of Minnesota Digital
  Conservancy, 2014. [Online]. Available:
  \url{https://hdl.handle.net/11299/163884}
\BIBentrySTDinterwordspacing

\bibitem{2023DRL-AoI-Long}
Y.~Long, J.~Zhuang, S.~Gong, B.~Gu, J.~Xu, and J.~Deng, ``Exploiting deep
  reinforcement learning for stochastic {AoI} minimization in
  multi-{UAV}-assisted wireless networks,'' in \emph{proc. IEEE WCNC}, Dubai,
  United Arab Emirates, Apr. 2024, pp. 1--6.

\bibitem{2024-AOI-LONG}
Y.~Long, S.~Zhao, S.~Gong, B.~Gu, D.~Niyato, and X.~Shen, ``{AoI}-aware sensing
  scheduling and trajectory optimization for multi-{UAV}-assisted wireless
  backscatter networks,'' \emph{IEEE Trans. Veh. Technol.}, pp. 1--16, May
  2024, doi:10.1109/TVT.2024.3402740.

\bibitem{2022UAVsensingZhu}
B.~Zhu, E.~Bedeer, H.~H. Nguyen, R.~Barton, and Z.~Gao, ``{UAV} trajectory
  planning for {AoI}-minimal data collection in {UAV}-aided {IoT} networks by
  transformer,'' \emph{IEEE Trans. Wireless Commun.}, vol.~22, no.~2, pp.
  1343--1358, Feb. 2023.

\bibitem{2024-UAV-AoI-Qi}
W.~Qi, C.~Yang, Q.~Song, Y.~Guan, L.~Guo, and A.~Jamalipour, ``Minimizing age
  of information for hybrid {UAV-RIS}-assisted vehicular networks,'' \emph{IEEE
  Internet Things J.}, vol.~11, no.~10, pp. 17\,886--17\,895, Feb. 2024.

\bibitem{2024-UAV-AoI-Chen}
B.~Chen, D.~Liu, J.~Zhang, and L.~Hanzo, ``Learning-aided {UAV}-cooperation
  reduces the age-of-information in wireless networks,'' \emph{IEEE Commun.
  Lett.}, vol.~28, no.~5, pp. 1053--1057, Feb. 2024.

\bibitem{2022-scheduling-UAV-samir}
M.~Samir, C.~Assi, S.~Sharafeddine, and A.~Ghrayeb, ``Online altitude control
  and scheduling policy for minimizing {AoI} in {UAV}-assisted {IoT} wireless
  networks,'' \emph{IEEE Trans. Mob. Comput.}, vol.~21, no.~7, pp. 2493--2505,
  Jul. 2022.

\bibitem{2024-UAV-scheduling-AoI-Sun}
Q.~Sun, J.~Niu, X.~Zhou, T.~Jin, and Y.~Li, ``{AoI} and data rate optimization
  in aerial {IRS}-assisted {IoT} networks,'' \emph{IEEE Internet Things J.},
  vol.~11, no.~4, pp. 6481--6493, Feb. 2024.

\bibitem{2022-UAV-semantic-Kang}
X.~Kang, B.~Song, J.~Guo, Z.~Qin, and F.~R. Yu, ``Task-oriented image
  transmission for scene classification in unmanned aerial systems,''
  \emph{IEEE Trans. Commun.}, vol.~70, no.~8, pp. 5181--5192, Aug. 2022.

\bibitem{2024-UAV-semantic-Yue}
P.~Yue, J.~Xin, Y.~Zhang, Y.~Lu, and M.~Shan, ``Semantic-driven autonomous
  visual navigation for unmanned aerial vehicles,'' \emph{IEEE Trans. Ind.
  Electron.}, pp. 1--11, Mar. 2024, doi:10.1109/TIE.2024.3363761.

\bibitem{2023-UAV-semantic-Kang}
J.~Kang, H.~Du, Z.~Li, Z.~Xiong, S.~Ma, D.~Niyato, and Y.~Li, ``Personalized
  saliency in task-oriented semantic communications: {Image} transmission and
  performance analysis,'' \emph{IEEE J. Sel. Areas Commun.}, vol.~41, no.~1,
  pp. 186--201, Jan. 2023.

\bibitem{2023-UAV-semantic-Wang}
L.~Wang, W.~Wu, F.~Tian, and H.~Hu, ``Intelligent resource allocation for
  {UAV}-enabled spectrum sharing semantic communication networks,'' in
  \emph{prof. IEEE ICCT}, Oct. 2023, pp. 1359--1363.

\bibitem{2023-AoI-semantic-Chen}
J.~Chen, S.~Yang, T.-T. Chan, and H.~Pan, ``Enhancing information freshness via
  knowledge graph-aided semantic communication,'' in \emph{proc. IEEE ICICN},
  Aug. 2023, pp. 155--160.

\bibitem{2024-aoi-semantic-Liew}
Z.~Q. Liew, M.~Xu, W.~Y.~B. Lim, Z.~Xiong, D.~Niyato, and D.~I. Kim,
  ``Mechanism design for semantic communication in {UAV}-assisted metaverse:
  {A} combinatorial auction approach,'' \emph{IEEE Trans. Veh. Technol.},
  vol.~73, no.~2, pp. 2236--2251, Feb. 2024.

\bibitem{2023-AoI-semantic-Meng}
S.~Meng, S.~Wu, A.~Li, and Q.~Zhang, ``Toward goal-oriented semantic
  communications: {AoII} analysis of coded status update system under {FBL}
  regime,'' \emph{IEEE J. Sel. Areas Inf. Theory}, vol.~4, pp. 718--733, Dec.
  2023.

\bibitem{2018multiUAV-tra-wu}
Q.~Wu, Y.~Zeng, and R.~Zhang, ``Joint trajectory and communication design for
  multi-{UAV} enabled wireless networks,'' \emph{IEEE Trans. Wireless Commun.},
  vol.~17, no.~3, pp. 2109--2121, Jan. 2018.

\bibitem{2019Zhang-TWC_UAV}
G.~Zhang, Q.~Wu, M.~Cui, and R.~Zhang, ``Securing {UAV} communications via
  joint trajectory and power control,'' \emph{IEEE Trans. Wireless Commun.},
  vol.~18, no.~2, pp. 1376--1389, Jan. 2019.

\bibitem{2023-uav-Li}
J.~Li, S.~Li, and C.~Xue, ``Resource optimization for multi {UAV} formation
  communication based on {DQSEnet},'' in \emph{prof. ICCCS}, Apr. 2023, pp.
  351--356.

\bibitem{2023-Yang-semantic-UAV-JSAC}
Z.~Yang, M.~Chen, Z.~Zhang, and C.~Huang, ``Energy efficient semantic
  communication over wireless networks with rate splitting,'' \emph{IEEE J.
  Sel. Areas Commun.}, vol.~41, no.~5, pp. 1484--1495, May 2023.

\bibitem{2023-semantic-Cang}
Y.~Cang, M.~Chen, Z.~Yang, Y.~Hu, Y.~Wang, C.~Huang, and Z.~Zhang, ``Online
  resource allocation for semantic-aware edge computing systems,'' \emph{IEEE
  Internet Things J.}, pp. 1--1, Oct. 2023, doi:10.1109/JIOT.2023.3325320.

\bibitem{2023-semantic-utility-wang}
Z.~Wang and J.~Zhao, ``Utility-oriented wireless communications for {6G}
  networks: {Semantic} information transfer for {IRS} aided vehicular
  metaverse,'' in \emph{proc. IEEE VTC-Spring}, Florence, Italy, Jun. 2023, pp.
  1--7.

\bibitem{2006-Lya-queue}
M.~Neely, ``Energy optimal control for time-varying wireless networks,''
  \emph{IEEE Trans. Inf. Theory}, vol.~52, no.~7, pp. 2915--2934, Jul. 2006.

\end{thebibliography}
\end{document}